\documentclass[prl,twocolumn,superscriptaddress,nofootinbib]{revtex4}
\usepackage[dvipdfmx,hiresbb]{graphicx}
\usepackage{color}
\usepackage{enumerate}
\usepackage{epsfig,subfigure}
\usepackage{amsmath,amssymb,latexsym}
\usepackage{ascmac}
\usepackage{bm}
\newtheorem{theorem}{{\it \bf Theorem}}
\newtheorem{property}{{\it \bf Property}}

\newtheorem{lemma}{{\it \bf Lemma}}
\newtheorem{fact}{{\it \bf Fact}}
\def\U#1{{\rm #1}}

\newcommand{\bra}[1]{\langle #1 |}
\newcommand{\ket}[1]{| #1 \rangle}

\newcommand{\expect}[1]{\left\langle #1 \right\rangle}

\begin{document}
\title{
  Information-theoretic security proof of differential-phase-shift quantum key distribution protocol based on complementarity
}
\author{Akihiro Mizutani\footnote{mizutani@qi.mp.es.osaka-u.ac.jp}}
\affiliation{Graduate School of Engineering Science, Osaka University,
  Toyonaka, Osaka 560-8531, Japan}
\author{Toshihiko Sasaki}
\affiliation{Photon Science Center, Graduate School of Engineering, 
  The University of Tokyo, Bunkyo-ku, Tokyo 113-8656, Japan}
\author{Go~Kato}
\affiliation{NTT Communication Science Laboratories, NTT Corporation, 3-1,
Morinosato Wakamiya Atsugi-Shi, Kanagawa, 243-0198, Japan}
\author{Yuki Takeuchi}
\affiliation{Graduate School of Engineering Science, Osaka University,
  Toyonaka, Osaka 560-8531, Japan}
\author{Kiyoshi Tamaki}
\affiliation{
Department of Intellectual Information Engineering, Faculty of Engineering, University of Toyama, Gofuku 3190, Toyama 930-8555, Japan}
\begin{abstract}
  We show the information-theoretic security proof of the differential-phase-shift (DPS) quantum key distribution (QKD) protocol
  based on the complementarity approach~[arXiv:0704.3661 (2007)].
  Our security proof provides a slightly better key generation rate compared to the one derived in the previous
  security proof in [arXiv:1208.1995 (2012)] that is based on the Shor-Preskill approach~
  [Phys. Rev. Lett. {\bf 85}, 441 (2000)]. 
  This improvement is obtained because the complementarity approach can employ
  more detailed information on Alice's sending state in estimating the leaked information to an eavesdropper.
  Moreover, we remove the necessity of the numerical calculation that was needed in the previous analysis
  to estimate the leaked information. This leads to an advantage that our security proof enables 
  us to evaluate the security of the DPS protocol with any block size. 
  This paper highlights one of the fundamental differences between the Shor-Preskill and the complementarity approaches.
\end{abstract}
\maketitle

\section{Introduction}
\label{sec:introduction}
Quantum key distribution (QKD) holds promise to achieve information-theoretically
secure communication between two distant parties (Alice and Bob) against any eavesdropper (Eve).
Since the first invention of the BB84 protocol~\cite{bb84}, many QKD protocols have been proposed so
far~\cite{e91,b92,six,sarg04,cow,PhysRevLett.89.037902,dps2003,CV2002}. 
Among them, the differential-phase-shift (DPS) QKD protocol~\cite{dps2003} has
been considered as one of the promising protocols for future implementation 
since this protocol can be rather simply implemented with a passive detection unit.
Recently, a field demonstration of the DPS protocol~\cite{tokyoQKD} has already been conducted, and the 
information-theoretic security proofs of the DPS protocol have been established when Alice employs a single-photon source
~\cite{PhysRevLett.103.170503} and a block-wise phase-randomized coherent light source~\cite{Kiyoshi2012dps}.

The previous security proof~\cite{Kiyoshi2012dps} with coherent light source is based on the {\it Shor-Preskill approach}
~\cite{PhysRevLett.85.441} in which 
Alice and Bob virtually extract a maximally-entangled state (MES) to show that they share a monogamy correlation. 
In order to extract an MES, Alice and Bob use some estimated information about the correlation between them.
Specifically, this information consists of the bit and phase error rates, where the phase error is defined by
fictitious erroneous outcomes when Alice and Bob would have measured their virtual qubits
in a basis conjugate to the basis for generating the key. 
Since the phase error rate cannot be directly obtained in the experiment, the estimation of this quantity
is a central issue in the security proof, and some security proofs have been conducted along this approach
~\cite{PhysRevLett.90.057902,PhysRevLett.90.167904,PhysRevLett.93.120501,
  PhysRevLett.94.040503,PhysRevA.73.010302,PhysRevLett.103.170503}.

Another approach for the security proof is the complementarity approach~\cite{Koashi2007}. 
In this approach, a complementary control of the actual protocol and a virtual protocol are considered, 
which Alice and Bob choose to execute, but cannot execute simultaneously.
The goal of the actual protocol is to agree on the bit values along the key generation basis, say the $X$ basis, while 
in the virtual protocol, Alice and Bob collaborate to create an eigenstate of the $Z$ basis (a complementary basis to the
$X$ basis) in Alice's side.
With these protocols, Koashi proved in~\cite{Koashi2007} that the necessary and
sufficient condition for the secure key distillation is to be able to execute whichever task was chosen. 
On one hand, once an MES is shared between Alice and Bob, they also accomplish the complementary task, which implies that 
the Shor-Preskill approach is included in the complementarity one.
On the other hand, the purpose of the complementarity approach is to create an eigenstate of the $Z$ basis at Alice's side, and 
therefore, we can employ some additional information, such as the one on Alice's sending state,
which may provide an advantage over the Shor-Preskill approach.

In this paper, we show that these two approaches indeed give a different resulting secret key rate of the
DPS QKD protocol by exploiting a property of pulses emitted by Alice.
More specifically, we adopt the complementarity approach for the security proof where we accommodate the intuition that 
it is difficult to extract information from a train of weak coherent pulses employed in the DPS protocol.
As a result, we show that the secure key rate based on the complementarity approach is 1.22 times
as high as the one based on the Shor-Preskill approach when the bit error rate is 2\%.
Moreover, we remove the necessity of the numerical calculation that was needed to evaluate the leaked information to Eve
in the previous analysis~\cite{Kiyoshi2012dps}, and we provide the closed formulas for the upper bounds on the
leaked information.
This leads to an advantage that our security proof enables us to evaluate the security of the DPS protocol with any block size.

This paper is organized as follows. 
In Sec.~\ref{sec:DPSQKD}, we introduce the DPS protocol including the assumptions on Alice and Bob's devices. 
In Sec.~\ref{sec:security}, we explain our security proof based on the complementarity approach.
In our security proof, we estimate the leaked information for each photon number emission separately, and
in Sec.~\ref{sec:formula}, we show the resulting upper bounds on the estimated leaked information up to the two-photon emission
events. 
In Sec.~\ref{sec:key}, we compare the resulting secret key rates based on the Shor-Preskill and the complementarity approaches,
and finally, we conclude our paper in Sec.~\ref{sec:conc}. 

\section{DPS QKD}
\label{sec:DPSQKD}
\begin{figure*}[t]
  \centering
\includegraphics[width=10cm]{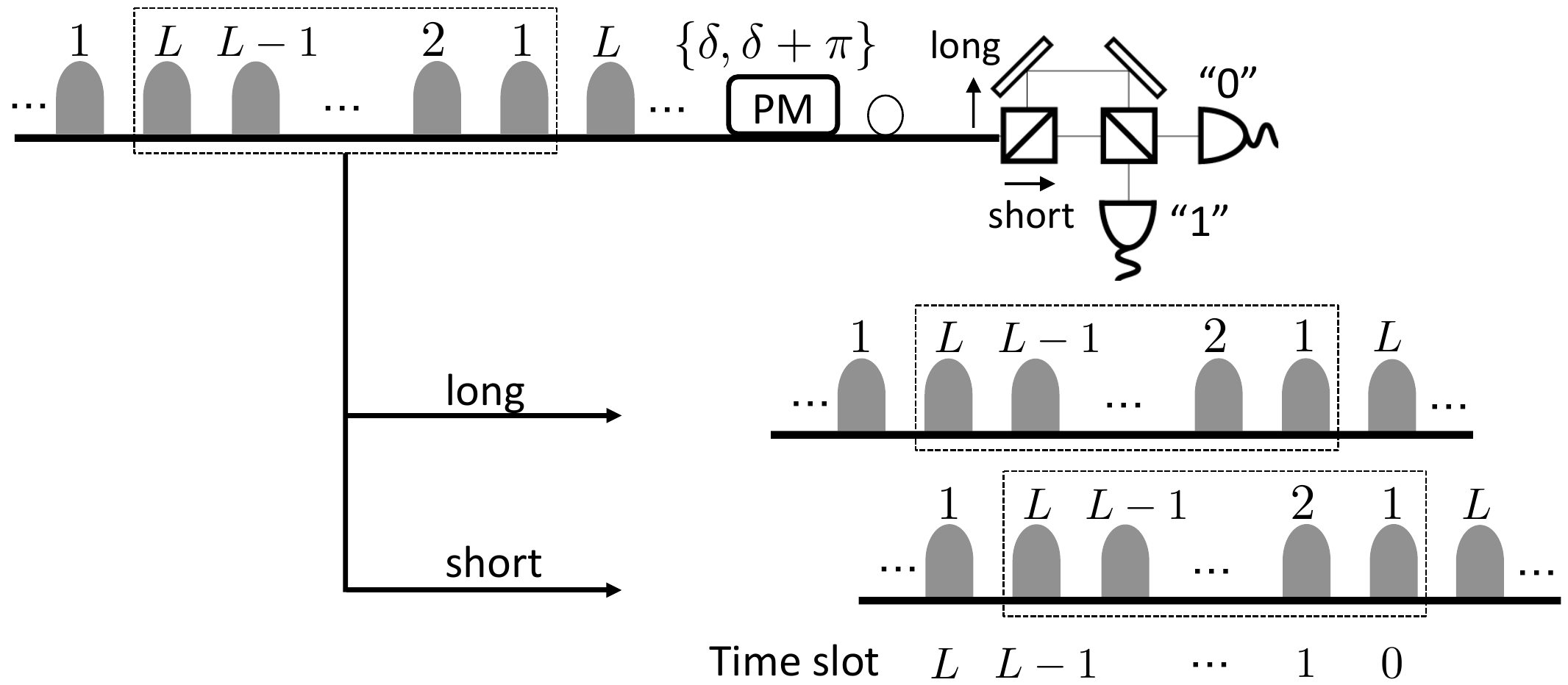}
\caption{
  Setup of the actual protocol. At Alice's site, pulse trains are generated by a laser source followed by the phase modulator~(PM)
  that
  randomly modulates a phase $\delta$ or $\delta+\pi$ with $\delta$ being randomly chosen from $[0,2\pi)$ for each block of $L$ pulses. 
    At Bob's site, each pulse train is fed to a one-bit delay Mach-Zehnder interferometer with two
    50:50 beam splitters. 
    The pulse trains leaving the interferometer are measured by two photon-number-resolving (PNR) 
    detectors corresponding to bit values ``0'' and ``1''.
    A successful detection event occurs if Bob detects a single-photon 
    in the only one time slot $j$ (with $1\leq j\leq L-1$), and detects the vacuum in all the other time slots
    including the $0^{\U{th}}$ and $L^{\U{th}}$ time slots. 
  }
  \label{fig:actual}
          \end{figure*}
We first describe the setup of the DPS protocol (see Fig.~\ref{fig:actual}),
and list up the assumptions we make on Alice and Bob's devices.
Note that the setup and the assumptions are exactly the same as those in~\cite{Kiyoshi2012dps}.

\subsection{Setup and assumptions}
\label{sec:setup}
Alice uses a laser source emitting coherent pulses and a phase modulator, and a train of $L$ $(L\ge3)$ pulses forms a block. 
Bob uses a one-bit delay interferometer with two 50:50 beam splitters and with
its delay being equal to the interval of the neighboring sending pulses. 
After the interferometer, the pulses are detected by two photon detectors corresponding to bit values of 0 and 1.
The $j^{\U{th}}$ ($1\le j\le L-1$) time slot is defined as an expected detection time
at Bob's detectors from the superposition of the $j^{\U{th}}$ and $(j+1)^{\U{th}}$ incoming pulses.
Also, the $0^{\U{th}}$ ($L^{\U{th}}$) time slot is defined as an expected detection time at Bob's detectors
from the superposition of the $1^{\U{st}}$ ($L^{\U{th}}$) incoming pulse and the $L^{\U{th}}$
incoming pulse in the previous block ($1^{\U{st}}$ incoming pulse in the next block). 

As for the assumptions on Alice's device,
we assume that (A 1) the phase modulator randomly modulates each relative phase between adjacent sending pulses by 0 or $\pi$.
Moreover, (A 2) the randomization of 
overall optical phase $\delta$ is done for each block of $L$ pulses. This means that the quantum 
state of the $L$ pulses is written as a classical mixture of the total photon number contained in the $L$ pulses. 
Besides, (A 3) we do not consider any side-channel in Alice's site. 

Regarding the assumptions on Bob's device,
we suppose that (B 1) Bob uses two photon-number-resolving (PNR) detectors, which can discriminate among the vacuum,
a single-photon and multiphoton. Also, we assume that (B 2) the detection efficiency is the same for both detectors. 
Finally, (B 3) we do not consider any side-channel in Bob's site. 

\subsection{Actual protocol}
\label{sec:protocol}
The actual protocol proceeds as follows.
In its description, $|\bm{\kappa}|$ denotes the length of a bit string $\bm{\kappa}$. 

(a 1) Alice generates a random $L$-bit sequence $s_1s_2...s_L$ and a random common phase shift $\delta\in[0,2\pi)$.
  For a random $L$-bit sequence $s_1s_2...s_L$, she sends the following coherent state (system $\mathcal{H}_S$)
  to Bob through a quantum channel
  \begin{align}
    \bigotimes^L_{i=1}\ket{e^{\U{i}\delta}(-1)^{s_i}\alpha}_{S,i},
\label{sendingstate}
    \end{align}
  where $\ket{\alpha}_{S,i}$
  represents the coherent state $\sum_ne^{-|\alpha|^2/2}\alpha^n\ket{n}_{S,i}/\sqrt{n!}$ of the $i^{\U{th}}$ pulse mode. 
  
  (a 2) Bob performs the interference measurement with two PNR detectors,
  and we call the {\it successful detection} event 
  if Bob detects one-photon in the $j^{\U{th}}$ time slot (with $1\leq j\leq L-1$), and detects the vacuum in all the other time slots
  including the $0^{\U{th}}$ and $L^{\U{th}}$ time slots. 
  If the successful detection occurs, the variable $j$ is set to the time slot, otherwise Bob sets $j=0$.
  If $j\neq0$, he obtains his raw bit $s\in\{0,1\}$ depending on which detector has reported a 
  detection at the $j^{\U{th}}$ time slot. Bob announces $j$ over an authenticated public channel.

  (a 3) If $j\neq0$, Alice calculates her raw key bit as $s_j\oplus s_{j+1}\in\{0,1\}$.

  (a 4) Alice and Bob repeat steps (a 1)-(a 3) $N$ times.

  (a 5) Alice defines a sifted key $\bm{\kappa}_A$ by concatenating the successful detection events ({\it i.e.,} $j\neq0$).

  (a 6) Bob defines a sifted key $\bm{\kappa}_B$ by concatenating the successful detection events ({\it i.e.,} $j\neq0$). 

  (a 7) Bob corrects the errors in his sifted key $\bm{\kappa}_B$ to make it coincide with $\bm{\kappa}_A$ by sacrificing
  $|\bm{\kappa}_A|f_{\U{EC}}$ bits of encrypted public communication from Alice by consuming the same length of the 
  pre-shared secret key.

  (a 8) Alice and Bob conduct privacy amplification by shortening their keys by $|\bm{\kappa}_A|f_{\U{PA}}$ to obtain the
  final keys.

  In this paper, we consider the asymptotic limit of the sifted key length ($N\rightarrow\infty$).
  In the experiments, the following parameters are observed:
  \begin{align}
Q:=\frac{|\bm{\kappa}_A|}{N},~~e^{(\U{b})}:=\frac{\U{wt}(\bm{\kappa}_A-\bm{\kappa}_B)}{|\bm{\kappa}_A|},
    \end{align}
  where the minus sign is a bit-by-bit modulo-2 subtraction and
  $\U{wt}(\bm{\kappa})$ denotes the weight (the number of 1's) in a bit string $\bm{\kappa}$. 
  In the asymptotic limit, $f_{\U{EC}}$ is given by a function of the bit error rate $e^{(\U{b})}$.
  The asymptotic key generation rate per sending pulse is given by~\cite{GLLP04,Koashi2009}
  \begin{align}
    G=
    [Q(1-f_{\U{EC}}(e^{(\U{b})})-f_{\U{PA}})]/L.\label{keyrate}
    \end{align}
  In this key generation formula,
  we omit for simplicity the random sampling procedure to estimate the bit error rate $e^{(\U{b})}$ because its cost is 
  negligible in the asymptotic limit. 

  In step (a 2),
  when the successful detection event ($j\neq0$) occurs, the state of the incoming $L$ pulses is expressed by the
  Hilbert space $\mathcal{H}_B$ 
  spanned by $L$ states, and we denote its orthonormal basis by $\{\ket{i}_B\}^L_{i=1}$,
  with $i$ representing the position of the single-photon before the first beam splitter. 
  Determination of the detected time slot $j$ and the bit value $s\in\{0,1\}$ is represented by a generalized measurement
  on the system $\mathcal{H}_B$.
  Let $\hat{\Pi}_{j,s}$ be the POVM elements for the bit value $s$ detected at the $j^{\U{th}}$ time slot (with $1\leq j\leq L-1$).
  Considering the action of the beam splitters, they are written as
  \begin{align}
    \hat{\Pi}_{j,s}=\hat{P}\Big(\frac{\sqrt{\kappa_j}\ket{j}_B+(-1)^s\sqrt{\kappa_{j+1}}\ket{j+1}_B}{\sqrt{2}}\Big)
    \label{bobpovm}
  \end{align}
  with $\kappa_1=\kappa_L=1$ and $\kappa_i=1/2$ (for $2\leq i\leq L-1$).
  Here we define $\hat{P}(\ket{\cdot})=\ket{\cdot}\bra{\cdot}$.
  For later discussions, we decompose $\hat{\Pi}_{j,s}$ into two consecutive measurements.
  The first one is a filter operation $\hat{F}_j:\mathcal{H}_B\to \mathcal{H}_{B_{\U{q}}}$ with $1\leq j\leq L-1$,
  which gives the outcome $j$ and leaves a qubit system $\mathcal{H}_{B_{\U{q}}}$.
  The second one measures the qubit in the $Z$ basis $\{\ket{0}_{B_{\U{q}}},\ket{1}_{B_{\U{q}}}\}$.
  By using the filter operation and the qubit measurement, $\hat{\Pi}_{j,s}$ can be decomposed as
\begin{align}
  \hat{\Pi}_{j,s}=\hat{F}^{\dagger}_{j}\hat{P}(\ket{s}_{B_{\U{q}}})\hat{F}_{j}
  \label{PiFsF}
  \end{align}
if we choose $\hat{F}_j$ as
\begin{align}
  \hat{F}_j=\sqrt{\kappa_j}\ket{-}_{B_{\U{q}}}{}_B\bra{j}+\sqrt{\kappa_{j+1}}\ket{+}_{B_{\U{q}}}{}_B\bra{j+1}.
  \label{filter}
\end{align}
Here, we define the $X$ basis state as $\ket{\pm}=(\ket{0}\pm\ket{1})/\sqrt{2}$.

\section{Security proof}
\label{sec:security}
            In this section, we prove the security of the protocol described in Sec.~\ref{sec:protocol},
            and determine the amount of privacy amplification $f_{\U{PA}}$ in the asymptotic limit.

  \subsection{Alternative protocol}
  \label{sec:alternative}
\begin{figure}[t]
\includegraphics[width=8cm]{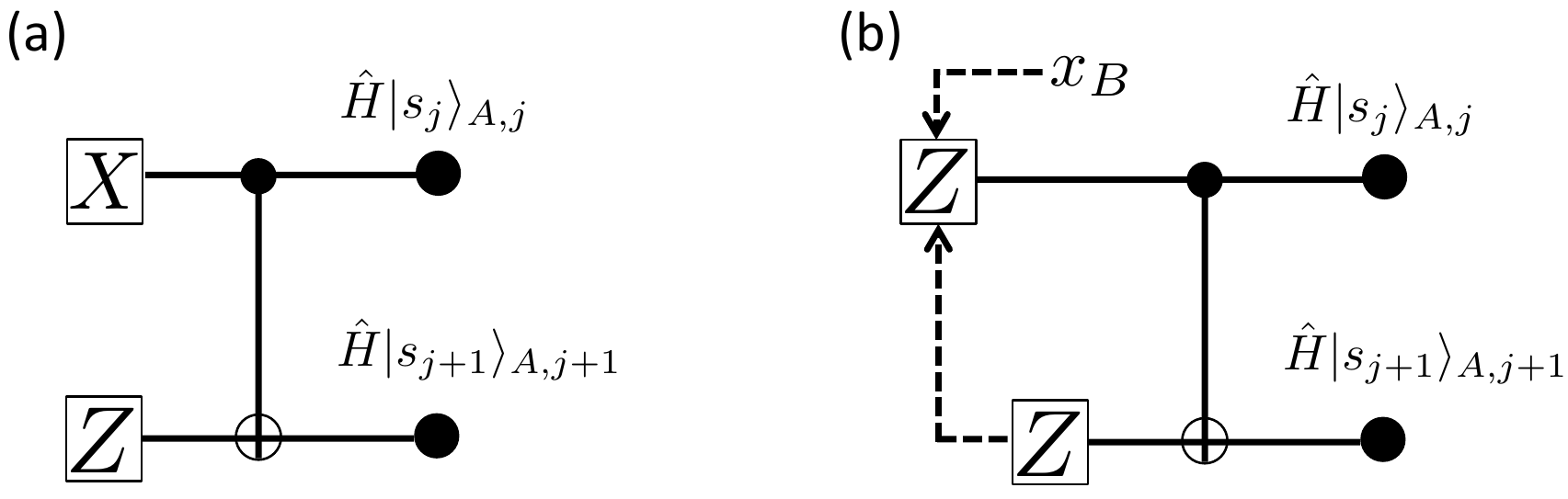}
\caption{
  (a) Alice's operation in the alternative protocol.
  She inputs the $j^{\U{th}}$ and $(j+1)^{\U{th}}$ qubits to the C-NOT gate with the $j^{\U{th}}$ one being the control and
  the $(j+1)^{\U{th}}$ one being the target. After that, the $(j+1)^{\U{th}}$ qubit is measured in the $Z$ basis, and
  the $j^{\U{th}}$ qubit is measured in the $X$ basis. 
  (b) Alice's procedure to estimate the outcome $z_j\in\{0,1\}$ of the complementary observable, that is, the $Z$ basis measurement 
  on the $j^{\U{th}}$ qubit. 
  After she performs the C-NOT gate, she measures her $(j+1)^{\U{th}}$ qubit in the $Z$ basis. 
  If $z_{j+1}=0~(1)$, she uses this information
  (this information and $x_B\in\{+,-\}$ in step (al 3$^\ast$)), and she predicts the outcome $z_j$. 
}
  \label{fig:circuit}
          \end{figure}
              To explain our security proof, we introduce an alternative protocol equivalent 
              to the actual one, which is designated to fulfill the following conditions. 
            
              {\it (i) The state of the optical pulses emitted by Alice and 
                the data processing for generating the final key are identical to the ones of the actual protocol. 
  
  (ii) Bob's measurement on receiving the $L$ pulses and his announcement of $j$ over an authenticated public channel 
    are identical to the actual protocol. }
              \\
              These two conditions ensure that Eve cannot change her attack 
              depending on which of the actual and the alternative protocols is conducted,
              resulting in the identical correlation between the final key and Eve's quantum system.
Hence, the final key in the actual protocol is secure in terms of the composable security~\cite{Canetti2001} if
the alternative protocol is secure against Eve's general attack.

As an alternative protocol, we consider an {\it entanglement-based} protocol in which Alice prepares
$L$ auxiliary qubits of system $\mathcal{H}_A$ located in Alice's site and $L$ coherent pulses of system $\mathcal{H}_S$ in state
\begin{align}
  &\ket{\Phi}_{C,A,S}=\sum^\infty_{\nu=0}\ket{\nu}_C\hat{\pi}_\nu\bigotimes^L_{i=1}\ket{\phi}_{A,S,i}
  \label{alterAlice}
\end{align}
with
\begin{align}
  &\ket{\phi}_{A,S,i}=\frac{1}{\sqrt{2}}\sum^1_{s_i=0}\hat{H}\ket{s_i}_{A,i}\ket{(-1)^{s_i}\alpha}_{S,i}. 
    \label{alterAlicei}
\end{align}
Here, $\hat{\pi}_\nu$ denotes the projection of the $L$ pulses onto the subspace where $\nu$ photons are contained in the $L$ pulses,
$\mathcal{H}_C$ is a system storing the information of the outcome of the projection,
and also we define the Hadamard operation as $\hat{H}=1/\sqrt{2}\sum_{x,y=0,1}(-1)^{xy}\ket{x}\bra{y}$.
To generate a raw key bit, Alice measures her auxiliary qubit in the $X$ basis.

Now, we introduce an alternative entanglement-based protocol that satisfies the above conditions (i) and (ii). 
A controlled-NOT (C-NOT) gate $\hat{U}^{(j)}_{\U{CNOT}}$ appearing in the protocol is defined on the $Z$ basis $\{\ket{0},\ket{1}\}$
by $\hat{U}^{(j)}_{\U{CNOT}}\ket{x}_{A,j}\ket{y}_{A,j+1}=\ket{x}_{A,j}\ket{x\oplus y}_{A,j+1}$ (with $x,y\in\{0,1\}$). 
   
              The protocol proceeds as follows.

              (al 1) Alice prepares the state $\ket{\Phi}_{C,A,S}$, 
              measures the system $\mathcal{H}_C$ to learn the total photon number $\nu$ in the $L$ pulses, 
              and sends the $L$ pulses to Bob through a quantum channel.
  
  (al 2) Bob receives the $L$ pulses and carries out the quantum nondemolition (QND)
  measurement to test if there is exactly 
  one photon in total from the $j=1^{\U{th}}$ to $j=(L-1)^{\U{th}}$ time slots and the vacuum in the 
  $j=0^{\U{th}}$ and $L^{\U{th}}$ time slots. If this test is passed, Bob performs the filter operation $\{\hat{F}_j\}^{L-1}_{j=1}$
  in Eq.~(\ref{filter}) to know in which time slot he detects a single-photon, and the variable $j$ is set to the time slot.
  Otherwise Bob sets $j=0$. Bob announces $j$ over an authenticated public channel. 
  If $j=0$, Alice and Bob skip steps (al 3) and (al 4) below. 
  
  (al 3) Bob measures his qubit $B_{\U{q}}$ in the $Z$ basis $\{\ket{0}_{B_{\U{q}}},\ket{1}_{B_{\U{q}}}\}$ and obtains his raw key bit $s$.

  (al 4-1) Alice applies the C-NOT gate on the $j^{\U{th}}$ qubit (control) and $(j+1)^{\U{th}}$ qubit (target)
  [see Fig.~\ref{fig:circuit}~(a)]. 

  (al 4-2) Alice measures the  $(j+1)^{\U{th}}$ qubit in the $Z$ basis $\{\ket{0}_{A,j+1},\ket{1}_{A,j+1}\}$ to obtain the outcome
  $z_{j+1}\in\{0,1\}$. 

  (al 4-3) Alice measures the $j^{\U{th}}$ qubit in the $X$ basis $\{\ket{+}_{A,j},\ket{-}_{A,j}\}$ and determines a raw key bit. 

  (al 5) Alice and Bob repeat steps (al 1)-(al 4) $N$ times. 
  
  (al 6)[=(a 5)] Alice defines a sifted key $\bm{\kappa}_A$ by concatenating the successful detection events ({\it i.e.,} $j\neq0$).

  (al 7)[=(a 6)] Bob defines a sifted key $\bm{\kappa}_B$ by concatenating the successful detection events ({\it i.e.,} $j\neq0$). 

  (al 8)[=(a 7)] Bob corrects the errors in his sifted key $\bm{\kappa}_B$ to make it coincide with $\bm{\kappa}_A$ by sacrificing
  $|\bm{\kappa}_A|f_{\U{EC}}$ bits of encrypted public communication from Alice by consuming the same length of the 
  pre-shared secret key.
\\

    This alternative protocol satisfies the above conditions (i) and (ii) because of the following reasons.
  First, as for (i), we show that Alice's procedure dictated in (i) is the same between both actual and alternative 
  protocols by modifying Alice's procedure in the alternative protocol. 
For this, since the outcome $z_{j+1}$ obtained in step (al 4-2) is neither announced nor used in determining the final key, we can 
omit this step. 
  Next, steps (al 4-1) and (al 4-3) are equivalently done by measuring all the $L$ qubits in the $X$ basis to obtain an 
  $L$-bit sequence $s_1s_2...s_L$ as the outcome, and then setting $s_{j}\oplus s_{j+1}$.
  Since the $X$-basis measurement on all the qubits does not require the knowledge of $j$ announced in step (al 2),
  we can consider that it is done in step (al 1).
  Then, using the relation
\begin{align}
{}_{A,j}\expect{\pm|\phi}_{A,S,j}=\frac{1}{\sqrt{2}}\ket{\pm\alpha}_{S,j},
\end{align}
we see that the random $L$-bit sequence $s_1s_2...s_L$ is obtained,
and we thus conclude that Alice's sending state is equivalent to Eq.~(\ref{sendingstate}). 
Note that
the total photon number measurement to obtain $\nu$ in step (al 1) makes the pulse train diagonalized in the Fock basis,
and with this measurement, the assumption (A 2) introduced in Sec.~\ref{sec:setup} is satisfied. 
Hence, the state of the optical pulses emitted by Alice and the data processing for generating the final key are identical to the 
ones of the actual protocol. 
 
  Next, we see that the alternative protocol satisfies the condition (ii). 
  In step (al 2), the QND measurement over the $j=0^{\U{th}}$ to $j=L^{\U{th}}$ time slots informs Bob
  if the successful detection occurs or not (namely, $j=0$ or $j\neq0$). 
  The probability of resulting $j\neq0$ is the same as the one in the actual protocol. 
  Moreover, if $j\neq0$, the probability of announcing $j$ (with $1\leq j\leq L-1$) is the same between both protocols
  since the following equation is satisfied from Eq.~(\ref{PiFsF}),
  \begin{align}
\sum^{1}_{s=0}\hat{\Pi}_{j,s}=\hat{F}^{\dagger}_j\hat{F}_j.
  \end{align}
  Hence, the information of $j$ announced by Bob in the alternative protocol is equivalent to the actual protocol. 

  Therefore, the alternative protocol satisfies the conditions (i) and (ii), which means that
  the security of the alternative protocol guarantees the security of the actual protocol.
            
  \subsection{Complementary task}
  \label{sec:complementarity}
  The next step is to determine the amount of privacy amplification $f_{\U{PA}}$, where we employ
  the argument of complementarity~\cite{Koashi2007}. 
  In this approach, we consider a {\it virtual measurement} that is complementary to the one to determine the 
  sifted key $\bm{\kappa}_A$.
  Recall that the measurement to obtain the sifted key is performed in step (al 4-3), and the complementary measurement means that
  the measurement basis ($X$ basis)
  in step (al 4-3) is replaced with the complementary basis (here we consider the $Z$ basis).
  In other words, the virtual measurement is described as the following step (al 4-3$^{\ast}$). 
  
  (al 4-3$^*$) Alice measures the $j^{\U{th}}$ qubit in the $Z$ basis $\{\ket{0}_{A,j},\ket{1}_{A,j}\}$
  and determines the outcome $z_j\in\{0,1\}$.

  While Bob, instead of aiming at learning $\bm{\kappa}_A$, tries to guess the value of the complementary observable $z_j$. 
  This suggests that step (al 3) is replaced with the following step (al 3$^*$). 

  (al 3$^*$) Bob measures his qubit $B_{\U{q}}$ in the $X$ basis $\{\ket{+}_{B_{\U{q}}},\ket{-}_{B_{\U{q}}}\}$ and obtains its outcome 
  $x_B\in\{+,-\}$. 
  
  In the complementarity argument~\cite{Koashi2007}, we need to quantify how well Alice successfully predicts the outcome
  $z_j$ of the measurement defined in step (al 4-3$^*$) with a help of Bob through quantum communication 
  (see Fig.~\ref{fig:circuit}~(b) for Alice's procedure to estimate $z_j$).
  For the quantification, we employ the phase error rate.
  Here, the phase error rate $e^{(\U{ph})}$ is defined as the probability that Alice fails her prediction on $z_j$,
    where the phase error rate is related to the amount of privacy amplification $Qf_{\U{PA}}$~\cite{Koashi2009}. 
  
  To accomplish the prediction on $z_j$, Alice employs three information that could help her prediction, which
  are listed below. 
\begin{itemize}
\item (I-1) $x_B$ obtained in step (al 3$^\ast$).
\item(I-2) $z_{j+1}\in\{0,1\}$ obtained in step (al 4-2). 
\item  (I-3) The intensity of Alice's sending state is weak.
  \end{itemize}
  The information (I-1) informs Alice of which of the $j^{\U{th}}$ or $(j+1)^{\U{th}}$
  original pulse contains a single-photon, which corresponds to $x_B=-$ or $x_B=+$, respectively. 
  The failure probability for this prediction, which we call the phase error rate, is related to the amount of privacy amplification
  (see Eq.~(\ref{fheph}) for the explicit formula). 

  As an introduction, we first consider the simplest case where Eve is absent in the quantum channel
  (we call this case the normal operation).
  In this case, if Bob successfully detects a single-photon at the $j^{\U{th}}$ time slot, and 
  depending on his $X$ basis measurement outcome $x_B=+$ or $x_B=-$ from the information (I-1),
  the state of the $j^{\U{th}}$ and $(j+1)^{\U{th}}$ Alice's systems is written as 
  \begin{align}
   &\ket{\psi_+}_{A,j,j+1}:= \ket{0}_{A,j}\ket{1}_{A,j+1}
  \end{align}
  and
    \begin{align}
   &\ket{\psi_-}_{A,j,j+1}:=\ket{1}_{A,j}\ket{0}_{A,j+1},
  \end{align}
    respectively. 
    After Alice performs the C-NOT gate in step (al 4-1), $\ket{\psi_+}_{A,j,j+1}$ and $\ket{\psi_-}_{A,j,j+1}$ are transformed to 
    \begin{align}
      &\hat{U}^{(j)}_{\U{CNOT}}\ket{\psi_+}_{A,j,j+1}=\ket{0}_{A,j}\ket{1}_{A,j+1}
                \label{Uni+}
    \end{align}
    and
        \begin{align}
      &\hat{U}^{(j)}_{\U{CNOT}}\ket{\psi_-}_{A,j,j+1}=\ket{1}_{A,j}\ket{1}_{A,j+1},
          \label{Uni-}
    \end{align}
        respectively. 
        In this case, $z_{j+1}$ obtained in step (al 4-2) is always 1, and Alice can
        predict $z_j$ as 0 or 1 without causing any error depending on Bob's outcome $x_B=+$ or $-$, respectively. 

        Moreover, by following the same arguments above,
        if the quantum channel is a linear lossy channel $\mathcal{N}_{\eta}: \mathcal{N}_{\eta}\ket{\alpha}=\ket{\eta\alpha}$,
        $z_{j+1}=0$ never occurs, and for $z_{j+1}=1$, Alice can also perfectly predict the outcome $z_j$. 
Summarizing two examples of the normal operation and the linear lossy channel case, 
if $z_{j+1}=1$ as (I-2), the following Alice's prediction succeeds with unit probability,
\begin{align}
  \U{if}~z_{j+1}=1
\rightarrow z_j=
\begin{cases}
  0~\U{if}~x_B=+\\
  1~\U{if}~x_B=-.
\end{cases}
  \label{zjMz=1}
\end{align}
Therefore, in general case (without assuming any channel model), if $z_{j+1}=1$, we suppose that
Alice takes the above strategy on her prediction. 

As we discussed above, the successful detection events never occur with 
$z_{j+1}=0$ in the normal operation and the linear lossy channel case.
Hence, if the successful detection occurs with $z_{j+1}=0$, we consider that Alice always predicts $z_j$ as 0
because before Alice sends the system $\mathcal{H}_S$ to Bob, $z_j=0$ is more likely to occur.
This tendency is rather remarkable as the intensity of the sending state becomes weaker. 
Mathematically, this is confirmed as 
\begin{align}
&||{}_{A,j}\bra{z_j}{}_{A,j+1}\bra{0}\hat{U}^{(j)}_{\U{CNOT}}\bigotimes_{k=j,j+1}\ket{\phi}_{A,S,k}||^2\notag\\
  &\propto\frac{1+\expect{-\alpha|\alpha}[\expect{-\alpha|\alpha}+(-1)^{z_j}2]}{2(1+\expect{-\alpha|\alpha}^2)}=:p(\alpha,z_j)
\end{align}
and $p(\alpha,0)/p(\alpha,1)=(\coth\alpha^2)^2\geq1$. 
Here, $||\cdot||$ denotes the trace norm. 

From this consideration and (I-3), in any channel model, we suppose that Alice takes the following strategy as
\begin{align}
  \U{if}~z_{j+1}=0\rightarrow z_j=0.
  \label{zjMz=0}
\end{align}
It is notable that if $z_{j+1}=0$, Alice does not use Bob's information $x_B$ [namely, (I-1)] to estimate the outcome $z_j$, and
her prediction is wrong when $z_j=1$. 

This prediction strategy highlights the difference between the complementarity approach and Shor-Preskill approach. 
Recall that in the Shor-Preskill approach~\cite{Kiyoshi2012dps}, the goal of Alice and Bob in an alternative protocol
is to generate the maximally entangled state (MES) of the two qubit systems $\mathcal{H}_{A,j}$ and $\mathcal{H}_{B_{\U{q}}}$. 
To generate the MES, the bit and phase error rates are needed, where the phase errors are defined as the instances where 
Alice's $Z$ basis measurement outcome in step (al 4-3$^*$) and Bob's $X$ basis measurement outcome in step (al 3$^*$)
are different.
Hence, the phase error is defined as the relation between $z_j$ and $x_B$, 
which means that it cannot be defined as an instance only by focusing on Alice's specific measurement outcome,
such as $z_j=1$. 
In fact, if we translate the construction of the phase error POVM in~\cite{Kiyoshi2012dps} 
to the complementarity argument, Alice's prediction for $z_{j+1}=0$ in~\cite{Kiyoshi2012dps} can be interpreted as follows: 
\begin{align}
  \U{if}~z_{j+1}=0
\rightarrow z_j=
\begin{cases}
  0~\U{with~probability~1/2}\\
  1~\U{with~probability~1/2}.
  \label{zj+10SP}
\end{cases}
\end{align}
This means that if $z_{j+1}=0$, Alice randomly selects $z_j$, 
which implies that the knowledge on Alice's sending states [namely, (I-3)] is not used. 

Therefore, the prediction in Eq.~(\ref{zjMz=0}) that uses a property of Alice's sending state highlights 
one of the unique features in the complementarity approach, and
this difference of the prediction
leads to a tighter upper bound on the phase error rate, which we show in Sec.~\ref{sec:formula}. 
Specifically, this difference will be reflected by all the differences in two curves 
in Figs.~\ref{fig:single}, \ref{fig:two} and \ref{fig:key}.

Our goal is to obtain the amount of $Qf_{\U{PA}}$ in Eq.~(\ref{keyrate}).
  For this, recall that the photon number measurement is conducted on the $L$ pulses in step (al 1),
  and therefore, $Qf_{\U{PA}}$ in Eq.~(\ref{keyrate}) can be expressed as a classical mixture of Fock state, that is, 
  \begin{align}
    Qf_{\U{PA}}=\sum^\infty_{\nu=0}Q^{(\nu)}f^{(\nu)}_{\U{PA}}.
    \label{PAnu}
  \end{align}
  Here, $Q^{(\nu)}$ is defined as $Q^{(\nu)}=\frac{|\bm{\kappa}^{(\nu)}_A|}{N}$
  with $\bm{\kappa}^{(\nu)}_A$ denoting the sifted key originating from the $\nu$-photon emission events, and
  $f^{(\nu)}_{\U{PA}}$ is the amount of privacy amplification for $\bm{\kappa}^{(\nu)}_A$.
  From~\cite{GLLP04,Koashi2009}, in the asymptotic limit, the amount of privacy amplification for $\bm{\kappa}^{(\nu)}_A$ is given by
  \begin{align}
    f^{(\nu)}_{\U{PA}}=h(e^{(\U{ph},\nu)}),
    \label{fheph}
    \end{align}
  where $h(x):=-x\log_2x-(1-x)\log_2(1-x)$ represents the binary entropy function, and
  $e^{(\U{ph},\nu)}$ is a phase error rate for the $\nu$-photon emission events.
\\

Having finished the explanation of the prediction on $z_j$, next we give an overview of the proof conducted in the following sections 
(from Sec.~\ref{sec:povm} to Sec.~\ref{sec:two}). 
The goal of our discussion is to obtain the upper bound on the amount of privacy amplification 
$f^{(\nu)}_{\U{PA}}=h(e^{(\U{ph},\nu)})$ in Eq.~(\ref{fheph}) with the bit error rate $e^{(\U{b},\nu)}$. 
To achieve this goal, in Sec.~\ref{sec:povm}, we first formulate the POVM elements for the bit and phase error events.
In Sec.~\ref{sec:const}, to discuss the security for each photon number emission event separately,
we introduce the projection operator $\hat{P}^{(\nu)}$ representing that the state of $L$ pulses is contained in the
$\nu$-photon subspace.
In Sec.~\ref{sec:relation}, we relate $e^{(\U{ph},\nu)}$ and $e^{(\U{b},\nu)}$, and we show that 
it suffices to calculate the quantity $\Omega^{(\nu)}(\lambda)$ in Eq.~(\ref{relationeph})
to upper-bound the phase error rate $e^{(\U{ph},\nu)}$. 
Then, in Sec.~\ref{sec:formula}, we explicitly derive the quantity $\Omega^{(\nu)}(\lambda)$ for $\nu=0,1,2$, and
obtain the upper bound on $e^{(\U{ph},\nu)}$ with the bit error rate $e^{(\U{b},\nu)}$ and $\Omega^{(\nu)}(\lambda)$. 
  
  \subsection{Bit and phase error POVMs}
  \label{sec:povm}
  In this subsection, we construct the POVMs for the phase and bit errors. 
The POVM elements for the phase error when the successful detection occurs at the $j^{\U{th}}$ (with $1\leq j\leq L-1$) time slot, 
which act on the systems $\mathcal{H}_{A,j}$, $\mathcal{H}_{A,j+1}$ (just before the C-NOT gate in Fig.~\ref{fig:circuit})
and $\mathcal{H}_B$, are given by
\begin{align}
  \hat{e}^{(\U{ph})}_{j}=&\hat{P}(\ket{1}_{A,j}\ket{1}_{A,j+1})\otimes\sum^1_{s=0}\kappa_{j+s}\hat{P}(\ket{j+s}_B)\notag\\
  +&\sum^1_{s=0}\hat{P}(\ket{s}_{A,j}\ket{\overline{s}}_{A,j+1})
  \otimes\kappa_{j+s}\hat{P}(\ket{j+s}_B),
  \label{ephj}
\end{align}
where we define $\bar{s}=s\oplus1$.
Here, the first and second terms correspond to the failure prediction on $z_j$ for $z_{j+1}=1$
[whose prediction strategy is given in  Eq.~(\ref{zjMz=1})]
and $z_{j+1}=0$ [whose prediction strategy is given in Eq.~(\ref{zjMz=0})], respectively. 
Next, we construct the POVM element for the bit error. 
A bit error is the instance where Alice's $X$ basis measurement on her $j^{\U{th}}$ auxiliary 
qubit in step (al 4-3) and the outcome of Bob's interference measurement defined in Eq.~(\ref{bobpovm}) are different. 
From this definition, the POVM elements for the bit error for the $j^{\U{th}}$ time slot, 
which act on the systems $\mathcal{H}_{A,j}$, $\mathcal{H}_{A,j+1}$ and $\mathcal{H}_B$, are given by 
\begin{align}
  \hat{e}^{(\U{b})}_j=\sum_{s,s'}\hat{P}(\hat{H}\ket{s}_{A,j})\hat{P}(\hat{H}\ket{s'}_{A,j+1})\hat{\Pi}_{j,s\oplus s'\oplus1}.
  \label{biterrorPOVM}
\end{align}
For simplicity of analysis, we introduce the unitary operator $\hat{U}$ defined by
\begin{align}
  \hat{U}\bigotimes^L_{i'=1}(\hat{H}\ket{s_{i'}}_{A,i'})\ket{i}_B=(-1)^{s_i}\bigotimes^L_{i'=1}(\hat{H}\ket{s_{i'}}_{A,i'})\ket{i}_B
  \label{unitary}
\end{align}
for $1\leq i\leq L$. 
By applying $\hat{U}$ to $\hat{e}^{(\U{ph})}_{j}$ in Eq.~(\ref{ephj}), we obtain
the following equation (see Appendix~\ref{sec:apA} for the derivation)
\begin{align}
  &\hat{U}\hat{e}^{(\U{ph})}_{j}\hat{U}^{\dagger}=\sum_{\bm{a}}\hat{P}(\ket{\bm{a}}_A)\otimes\notag\\
  &\Big[\kappa_j
    \delta_{a_{j+1},1}\hat{P}(\ket{j}_B)+\kappa_{j+1}\delta_{a_j,1}\hat{P}(\ket{j+1}_B)\Big],
  \label{unieph}
\end{align}
where we denote Alice's $Z$ basis states as $\ket{\bm{a}}_A:=\ket{a_1}_{A,1}\ket{a_2}_{A,2}...\ket{a_L}_{A,L}$
with $\bm{a}:=a_1a_2...a_L$ ($a_i\in\{0,1\}$). 
Also, by applying unitary $\hat{U}$ to Eq.~(\ref{biterrorPOVM}), we have~\cite{Kiyoshi2012dps}
\begin{align}
\hat{U}\hat{e}^{(\U{b})}_j\hat{U}^{\dagger}=\hat{\Pi}_{j,1}.
  \end{align}
Then, by taking a sum over all the time slots, we obtain the operators for the phase and bit errors as
\begin{align}
  \hat{e}^{(\U{ph})}=\sum^{L-1}_{j=1}\hat{e}^{(\U{ph})}_{j},~~\hat{e}^{(\U{b})}=\sum^{L-1}_{j=1}\hat{e}^{(\U{b})}_j.\label{bitphase}
  \end{align}
When the state of Alice and Bob's quantum systems $\mathcal{H}_A$ and $\mathcal{H}_B$ just after the 
successful QND measurement ($j\neq0$) at step (al 2) is
$\hat{\rho}_{AB}$, the probability of having a bit error in the extracted qubit pair of systems 
$\mathcal{H}_{A_j}$ and $\mathcal{H}_{B_{\U{q}}}$ is given by $\U{tr}(\hat{\rho}_{AB}\hat{e}^{(\U{b})})$, and the probability of
having a phase error is given by $\U{tr}(\hat{\rho}_{AB}\hat{e}^{(\U{ph})})$.

By appling $\hat{U}$ to $\hat{e}^{(\U{ph})}$ and  $\hat{e}^{(\U{b})}$, these error operators are concisely written as follows. 
\begin{align}
  \hat{U}\hat{e}^{(\U{ph})}\hat{U}^{\dagger}=\sum_{\bm{a}}\hat{P}(\ket{{\bm{a}}}_A)\otimes\hat{\Pi}^{\U{(ph)}}_{{\bm{a}}},
 \label{UephU}
  \end{align}
where we define $\hat{\Pi}^{\U{(ph)}}_{{\bm{a}}}$ as
\begin{align}
  \hat{\Pi}^{\U{(ph)}}_{{\bm{a}}}=&\delta_{a_2,1}\hat{P}(\ket{1}_B)
  +\sum^{L-1}_{i=2}\frac{\delta_{a_{i-1},1}+\delta_{a_{i+1},1}}{2}\hat{P}(\ket{i}_B)\notag\\
  &+\delta_{a_{L-1},1}\hat{P}(\ket{L}_B),
  \label{PiUph}
\end{align}
and
\begin{align}
  \hat{U}\hat{e}^{(\U{b})}\hat{U}^{\dagger}&=\hat{I}_A\otimes\hat{\Pi}
  \label{UeU}
\end{align}
with
\begin{align}
\hat{\Pi}=\sum^{L-1}_{j=1}\hat{\Pi}_{j,1}.\label{Pi}
\end{align}
Here, $\hat{\Pi}$ is a tridiagonal symmetric matrix, and the matrix elements are given by
\begin{align}
  {}_B\bra{i}\hat{\Pi}\ket{i}_B&=1/2&(\U{for}~1\leq i\leq L),\notag\\
  {}_B\bra{i}\hat{\Pi}\ket{i+1}_B&=-1/(2\sqrt{2})&(\U{for}~i=1, L-1),\notag\\
  {}_B\bra{i}\hat{\Pi}\ket{i+1}_B&=-1/4&(\U{for}~2\leq i\leq L-2).
  \end{align}

\subsection{Constraints on Alice's auxiliary qubit system $\mathcal{H}_A$}
\label{sec:const}
Here, we constrain Alice's auxiliary qubit system $\mathcal{H}_A$ by using the knowledge of the
total photon number $\nu$ contained in the $L$ pulses. 
For this, we first rewrite Eq.~(\ref{alterAlice}) as
 \begin{align}
   \ket{\Phi}=2^{-L}\sum_{{\bm{a}}}\ket{{\bm{a}}}_A\sum_{\nu}\ket{\nu}_C\hat{\pi}_\nu
   \bigotimes^L_{i=1}(\ket{\alpha}_{i}+(-1)^{a_i}\ket{-\alpha}_i),
    \label{alterAlicecat}
  \end{align}
and from this equation, we see that if the total photon number is $\nu$, 
$\U{wt}({\bm{a}})\leq \nu$ is satisfied since $\ket{\alpha}-\ket{-\alpha}$ contains at least one photon. 
Also, since $\ket{\alpha}+(-)\ket{-\alpha}$ contains even (odd) number of photons, 
we also have that the parity of $\nu$ and $\U{wt}({\bm{a}})$ are the same, that is, $(-1)^{\nu}=(-1)^{\U{wt}({\bm{a}})}$. 
Therefore, after the successful detection event ($j\neq0$) occurs and the system $\mathcal{H}_C$ reveals a photon number $\nu$,
the state of Alice and Bob's systems are contained in the range projection operator $\hat{P}^{(\nu)}$ with
\begin{align}
\hat{P}^{(\nu)}:=\sum_{{\bm{a}}:\U{wt}({\bm{a}})=\nu,\nu-2,\nu-4...}\sum^L_{i=1}\hat{P}(\ket{{\bm{a}}}_A\ket{i}_B).
  \end{align}
Through the unitary $\hat{U}$ in Eq.~(\ref{unitary}), we have~\cite{Kiyoshi2012dps}
\begin{align}
  \hat{U}\hat{P}^{(\nu)}\hat{U}^{\dagger}&=
  \sum_{{\bm{a}}:\U{wt}({\bm{a}})=\nu-1,\nu-3...}\hat{P}(\ket{{\bm{a}}}_A)\otimes \hat{I}_B\notag\\
  &+\sum_{{\bm{a}}:\U{wt}({\bm{a}})=\nu+1}\hat{P}(\ket{{\bm{a}}}_A)\otimes\hat{P}_{{\bm{a}}}
  \label{UPnuU}
  \end{align}
with
\begin{align}
  \hat{P}_{\bm{a}}:=\sum^L_{i=1}\hat{P}(\ket{i}_B)\delta_{a_i,1}.
  \label{Pbma}
 \end{align}
Note that Eq.~(\ref{UPnuU}) can be derived by using Eq.~(\ref{unitaryph}) in Appendix~\ref{sec:apA}.

\subsection{Relation between the bit and phase errors}
\label{sec:relation}
In this subsection, we derive the upper bound on the phase error rate for $e^{(\U{ph},\nu)}$ for the $\nu$-photon emission events
by using the bit error rate $e^{(\U{b},\nu)}$ originating from the $\nu$-photon emission events. 
To obtain this, we consider deriving the largest eigenvalue $\Omega^{(\nu)}(\lambda)$ of the operator 
\begin{align}
  \hat{P}^{(\nu)}(\hat{e}^{(\U{ph})}-\lambda\hat{e}^{(\U{b})})\hat{P}^{(\nu)}=\hat{e}^{(\U{ph},\nu)}-\lambda\hat{e}^{(\U{b},\nu)}
  \label{Pnuephe}
\end{align}
with $0<\lambda<\infty$.
Here, we define the POVM elements regarding the phase and bit errors for the $\nu$-photon emission events as 
$\hat{e}^{(\U{ph},\nu)}:=\hat{P}^{(\nu)}\hat{e}^{(\U{ph})}\hat{P}^{(\nu)}$ and
$\hat{e}^{(\U{b},\nu)}:=\hat{P}^{(\nu)}\hat{e}^{(\U{b})}\hat{P}^{(\nu)}$, respectively. 
Once we obtain $\Omega^{(\nu)}(\lambda)$, we can bound the phase error rate for the $\nu$-photon emission events as
\begin{align}
  e^{(\U{ph},\nu)}\leq\lambda e^{(\U{b},\nu)}+\Omega^{(\nu)}(\lambda),
  \label{relationeph}
\end{align}
and hence to derive $\Omega^{(\nu)}(\lambda)$ is vital for determining the key rate. 
Since the unitary operator does not change the eigenvalues, $\Omega^{(\nu)}(\lambda)$ is also the largest eigenvalue of the
operator~\cite{Kiyoshi2012dps}. 
\begin{align}
  &\hat{U}\hat{P}^{(\nu)}\hat{U}^{\dagger}(\hat{U}\hat{e}^{(\U{ph})}\hat{U}^{\dagger}-\lambda\hat{U}\hat{e}^{(\U{b})}\hat{U}^{\dagger})
  \hat{U}\hat{P}^{(\nu)}\hat{U}^{\dagger}\notag\\
  &=\sum_{{\bm{a}}:\U{wt}({\bm{a}})=\nu-1,\nu-3...}\hat{P}(\ket{{\bm{a}}}_A)\otimes(\hat{\Pi}^{\U{(ph)}}_{{\bm{a}}}-\lambda\hat{\Pi})\notag\\
  &+\sum_{{\bm{a}}:\U{wt}({\bm{a}})
    =\nu+1}\hat{P}(\ket{{\bm{a}}}_A)\otimes \hat{P}_{\bm{a}}(\hat{\Pi}^{\U{(ph)}}_{{\bm{a}}}-\lambda\hat{\Pi})\hat{P}_{\bm{a}},
  \label{Ueph-eU}
\end{align}
where Eqs.~(\ref{UephU}), (\ref{UeU}) and (\ref{UPnuU}) are used. 
To obtain an upper bound on Eq.~(\ref{Ueph-eU}), we use a fact that Eq.~(\ref{Ueph-eU}) is a direct sum of
$\hat{\Pi}^{\U{(ph)}}_{{\bm{a}}}-\lambda\hat{\Pi}$ with different ${\bm{a}}$ of $\U{wt}({\bm{a}})=\nu-1,\nu-3...$ and
$\hat{P}_{\bm{a}}(\hat{\Pi}^{\U{(ph)}}_{{\bm{a}}}-\lambda\hat{\Pi})\hat{P}_{\bm{a}}$ with ${\bm{a}}$ of $\U{wt}({\bm{a}})=\nu+1$.
Since 
$\max_{\bm{a}}(\hat{\Pi}^{\U{(ph)}}_{{\bm{a}}}-\lambda\hat{\Pi})\geq
\max_{\bm{a'}}(\hat{\Pi}^{\U{(ph)}}_{{\bm{a'}}}-\lambda\hat{\Pi})$ 
holds for any ${\bm{a}}$ and ${\bm{a'}}$ with $\U{wt}({\bm{a}})\geq \U{wt}({\bm{a'}})$, we only need to consider
$\hat{\Pi}^{\U{(ph)}}_{\bm{a}}-\lambda\hat{\Pi}$ with $\U{wt}({\bm{a}})=\nu-1$. 
We thus conclude that $\Omega^{(\nu)}(\lambda)$ is the larger of the two numbers
$\Omega^{(\nu)}_-(\lambda)$ and $\Omega^{(\nu)}_+(\lambda)$ defined as follows; 
$\Omega^{(\nu)}_-(\lambda)$ is the largest eigenvalue of the operator
\begin{align}
\{\hat{\Pi}^{\U{(ph)}}_{\bm{a}}-\lambda\hat{\Pi}~|~\U{wt}({\bm{a}})=\nu-1\},
  \label{omega-} 
\end{align}
and $\Omega^{(\nu)}_+(\lambda)$, which is the largest eigenvalue of the operator
\begin{align}
  \{\hat{P}_{\bm{a}}(\hat{\Pi}^{\U{(ph)}}_{\bm{a}}-\lambda\hat{\Pi})\hat{P}_{\bm{a}}~|~\U{wt}({\bm{a}})=\nu+1\}.
  \label{omega+}
  \end{align}
Recall that $\hat{P}_{\bm{a}}$ is defined in Eq.~(\ref{Pbma}),
and $\hat{P}_{\bm{a}}(\hat{\Pi}^{\U{(ph)}}_{\bm{a}}-\lambda\hat{\Pi})\hat{P}_{\bm{a}}$ is contained
in the subspace $\{\ket{i}_B\}$ with $a_i=1$.

\section{Explicit relations between the bit and phase error rates}
\label{sec:formula}

\begin{figure}[t]
\includegraphics[width=6cm]{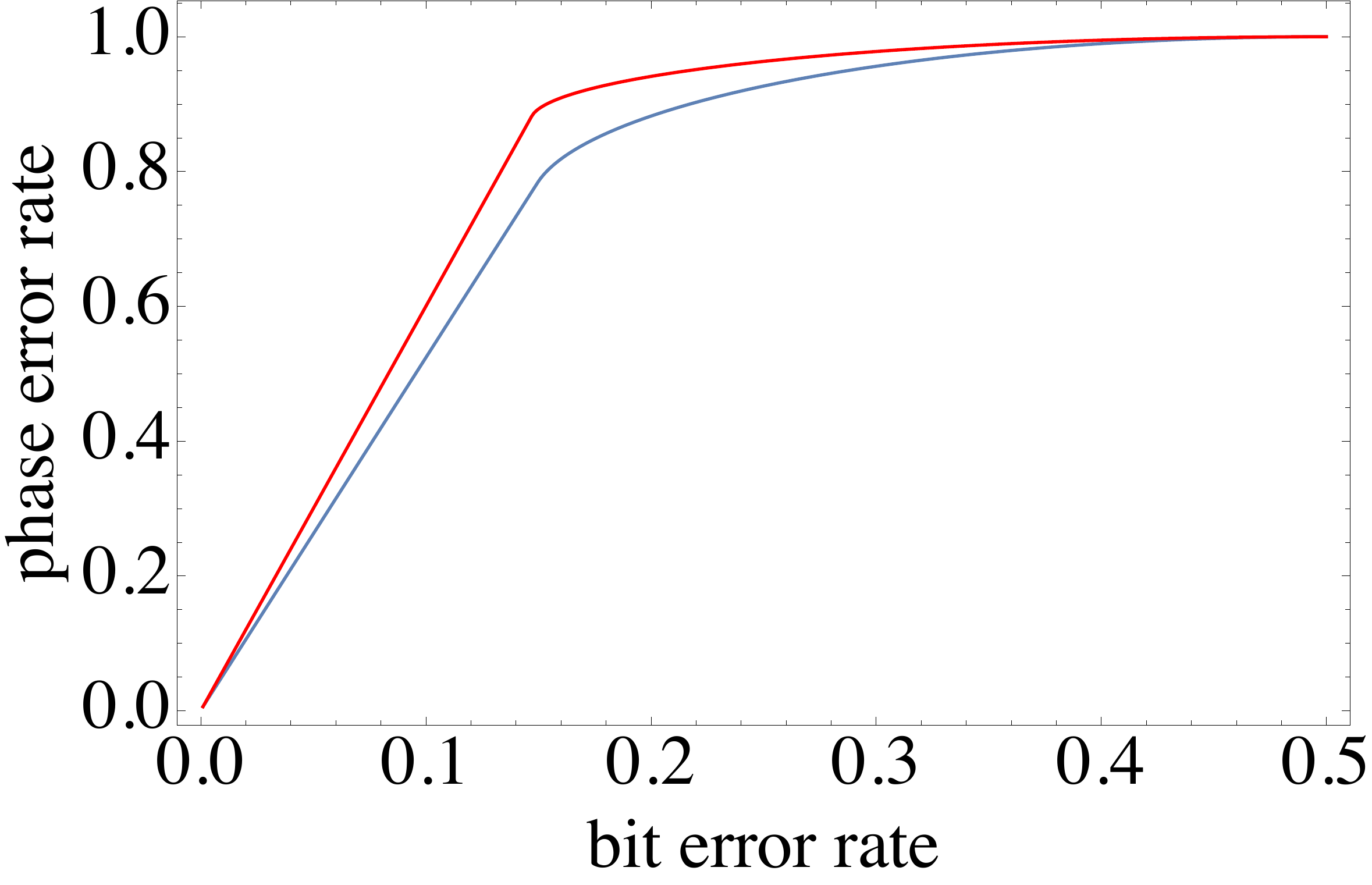}
\caption{
  Color online: 
  The upper bound on the phase error rates $e^{(\U{ph},1)}$ as a function of $e^{(\U{b},1)}$. 
  The red curve is based on the previous Shor-Preskill analysis~\cite{Kiyoshi2012dps} 
  and the blue one is our analysis based on the complementarity.
  Note that these bounds are independent of the block length $L$. 
}
\label{fig:single}
\end{figure}

Here, we derive the explicit relations between $e^{(\U{b},\nu)}$ and $e^{(\U{ph},\nu)}$ by evaluating $\Omega^{(\nu)}(\lambda)$. 
For simplicity, we consider extracting the secret key up to the two-photon emission events ($\nu=0,1$ and 2),
and for more than two-photon emission events, we pessimistically 
assume that Eve has perfect knowledge on the sifted keys. 
Therefore, it is sufficient to provide the relationship for $\nu=0,1$, and 2.

\subsection{Zero-photon part}
\label{sec:zero}
First, we discuss the case for $\nu=0$ when Alice emits zero-photon.
Since $\Omega^{(0)}_-(\lambda)$ has no candidates for $\nu=0$,
$\Omega^{(0)}(\lambda)=\Omega^{(0)}_+(\lambda)$.
For $\Omega^{(0)}_+(\lambda)$, regardless of $i$ such that $a_i=1$, we have
\begin{align}
  \Omega^{(0)}(\lambda)={}_B\bra{i}(\hat{\Pi}^{(\U{ph})}_{\bm{a}}-\lambda\hat{\Pi})\ket{i}_B=-\lambda/2,
\end{align}
which corresponds to the single fixed point $(e^{(\U{b},0)},e^{(\U{ph},0)})=(1/2,0)$. 

\subsection{Single-photon part}
\label{sec:single}
Next, we derive a relation between $e^{(\U{b},1)}$ and $e^{(\U{ph},1)}$. 
First, as for $\Omega^{(1)}_-(\lambda)$, since $\hat{\Pi}^{\U{(ph)}}_{\bm{a}}=0$ from Eq.~(\ref{omega-}),
$\Omega^{(1)}_-(\lambda)$ is a largest eigenvalue of $-\lambda\hat{\Pi}$, which is zero
\footnote{Note that
  ${}_B\bra{\psi}\hat{\Pi}\ket{\psi}_B=0$ is achieved only for the state 
$\ket{\psi}_B=[\sum^{L-1}_{i=2}\ket{i}_B+(\ket{1}_B+\ket{L}_B)/\sqrt{2}]/\sqrt{L-1}$.}. 
Next, for the derivation of $\Omega^{(1)}_+(\lambda)$, since $\U{wt}({\bm{a}})=2$ from its definition in Eq.~(\ref{omega+}), 
there are $\binom L2$ patterns to choose $i$ and $j$ such that $a_i=a_j=1$ holds. 
In order to find the pair $(i,j)$ to achieve the largest eigenvalue of Eq.~(\ref{omega+}), we use the following fact
(see \cite{Yuki2017} for its proof). 
\begin{fact}
  \label{fact1}
  Given two $n\times n$ real matrices $A=(A_{i,j})_{i,j}$ and $\tilde{A}=(\tilde{A}_{i,j})_{i,j}$,
  whose off-diagonal elements are non-negative, their largest eigenvalues 
  (respectively denoted by $\Lambda_A$ and $\Lambda_{\tilde{A}}$) have a relation
\begin{align}
  \Lambda_A\geq\Lambda_{\tilde{A}}
\end{align}
  if $A_{i,j}\geq\tilde{A}_{i,j}$ holds for any $i$ and $j$ $(1\leq i,j\leq n$).
\end{fact}
Thanks to this fact and Eqs.~(\ref{PiUph}) and (\ref{Pi}), 
we find that the largest eigenvalue $\Omega^{(1)}_+(\lambda)$ is achieved on the subspace 
$\{\ket{1}_B,\ket{2}_B\}$ [namely, $(i,j)=(1,2)$], and by calculating the largest eigenvalue of
$\hat{P}_{\bm{a}}(\hat{\Pi}^{\U{(ph)}}_{\bm{a}}-\lambda\hat{\Pi})\hat{P}_{\bm{a}}$ with $a_1=a_2=1$,
we have $\Omega^{(1)}_+(\lambda)$ in Lemma~\ref{lemma1}.
\begin{lemma}
  \label{lemma1}
$\Omega^{(1)}_+(\lambda)$ is given by
  \begin{align}
\Omega^{(1)}_+(\lambda)=(3-2\lambda+\sqrt{1+2\lambda^2})/4,
  \end{align}
  which is non-negative if $\lambda\leq (3+\sqrt{5})$. 
\end{lemma}
Then, by combining the results of $\Omega^{(1)}_\pm(\lambda)$,
$\Omega^{(1)}(\lambda)=\max\{\Omega^{(1)}_+(\lambda),\Omega^{(1)}_-(\lambda)\}$ is given by
\begin{align}
  \Omega^{(1)}(\lambda)=
  \begin{cases}
    0&(\lambda> 3+\sqrt{5})\\
    (3-2\lambda+\sqrt{1+2\lambda^2})/4 &(\lambda\leq 3+\sqrt{5}).
    \label{omega1}
    \end{cases}
\end{align}
Then, from Eq.~(\ref{omega1}),
an upper bound on the phase error rate for the single-photon emission events is given in the Theorem~\ref{theorem1}
(see Appendix~\ref{sec:appTheorem} for its proof). 
\begin{theorem}
  \label{theorem1}
  The upper bound on $e^{(\U{ph},1)}$ is given by
  \begin{align}
e^{(\U{ph},1)}\leq (3+\sqrt{5})e^{(\U{b},1)} 
  \end{align}
  if $0\leq e^{(\U{b},1)}\leq (10-3\sqrt{5})/22$ and 
  \begin{align}
    e^{(\U{ph},1)}\leq \inf_{0<\lambda <3+\sqrt{5}}\{\lambda e^{(\U{b},1)}+(3-2\lambda+\sqrt{1+2\lambda^2})/4\}
    \label{nu1high}
  \end{align}
if $(10-3\sqrt{5})/22<e^{(\U{b},1)}$.
\end{theorem}
    
In Fig.~\ref{fig:single}, we plot the resulting relation on ($e^{(\U{b},1)},e^{(\U{ph},1)})$ (see the blue curve). 
For comparison, we compare our result and the previous work~\cite{Kiyoshi2012dps} (see the red curve),
and we find that our security proof gives a tighter bound on $e^{(\U{ph},1)}$.

\subsection{Two-photon part}
\label{sec:two}
\begin{figure}[t]
\includegraphics[width=6cm]{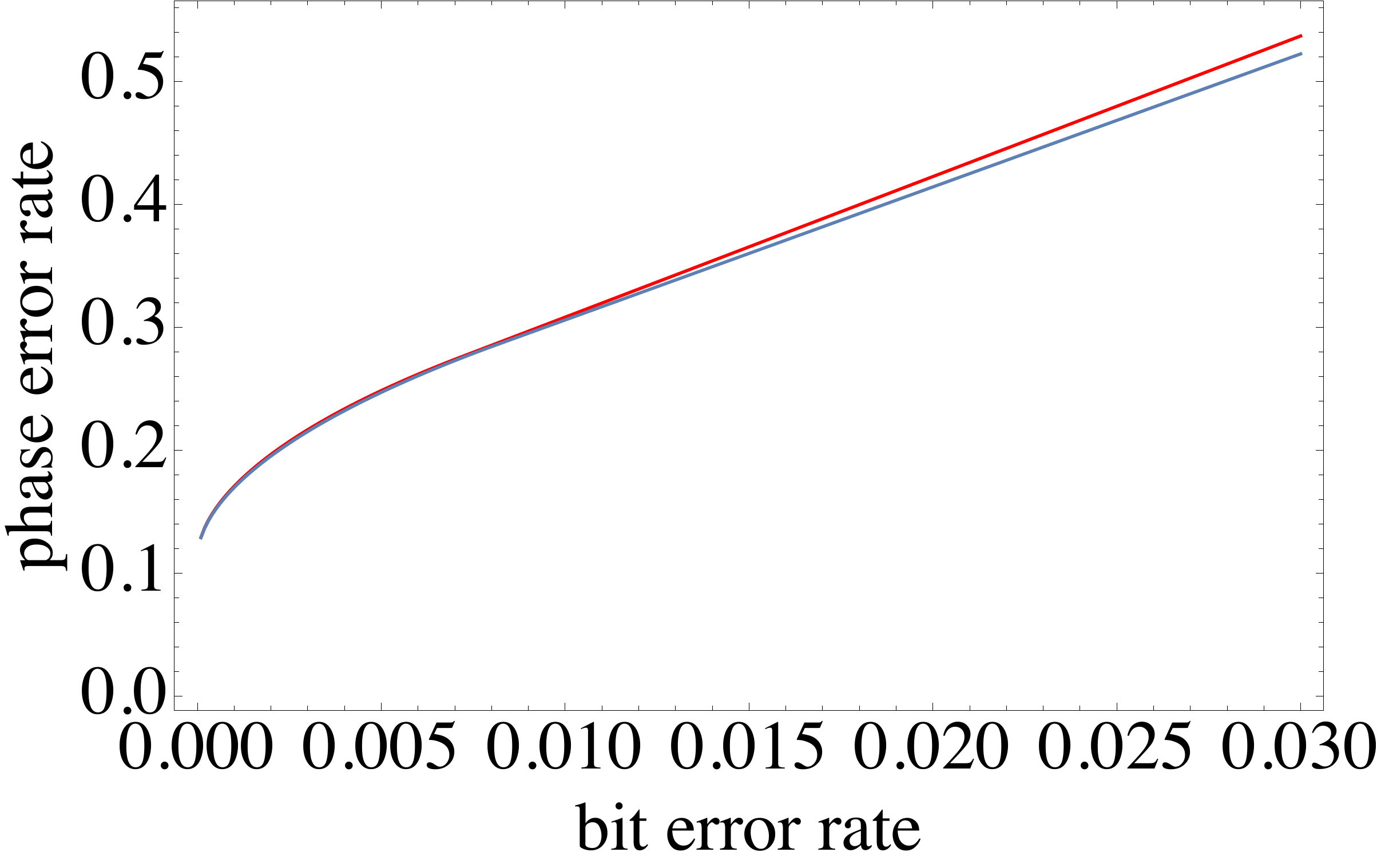}
\caption{
  Color online: 
  The upper bound on the phase error rates $e^{(\U{ph},2)}$ as a function of $e^{(\U{b},2)}$ with $L=10$. 
  The red curve is based on the previous Shor-Preskill analysis and the blue one is our analysis based on the complementarity.
}
\label{fig:two}
\end{figure}
Here, we derive an upper bound on $e^{(\U{ph},2)}$.
The evaluation of $\Omega^{(2)}(\lambda)=\max\{\Omega^{(2)}_+(\lambda),\Omega^{(2)}_-(\lambda)\}$ involves
the calculation of the largest eigenvalues $\Omega_+^{(2)}(\lambda)$ and $\Omega_-^{(2)}(\lambda)$. 
As for $\Omega_+^{(2)}(\lambda)$, since $\U{wt}({\bm{a}})=\nu+1=3$ from Eq.~(\ref{omega+}), 
we need to choose three indexes $1\leq i<j<k\leq L$ such that $a_i=a_j=a_k=1$ holds. 
There are $\binom L3$ patterns for the choices, and we need to find out the pair ($i,j,k$) that achieves 
the largest eigenvalue of Eq.~(\ref{omega+}). 
In so doing, the previous analysis~\cite{Kiyoshi2012dps} relies on a numerical method that compares each of
all the largest eigenvalues of Eq.~(\ref{omega+}) with different ($i,j,k$), which becomes complicated 
as $L$ becomes larger. 
On the other hand, we do not rely on a numerical method and derive $\Omega_+^{(2)}(\lambda)$ in a closed form as will be shown in
Theorem~\ref{lemma2}, which holds for any $L$ of $L\geq3$. 
This can be accomplished by using the fact \ref{fact1}, and as a result, 
we find that the largest eigenvalue $\Omega^{(2)}_+(\lambda)$ is obtained on the subspace 
$\{\ket{1}_B,\ket{2}_B,\ket{3}_B\}$ [namely, $(i,j,k)=(1,2,3)$]. 
By calculating the largest eigenvalue of 
$\hat{P}_{\bm{a}}(\hat{\Pi}^{\U{(ph)}}_{\bm{a}}-\lambda\hat{\Pi})\hat{P}_{\bm{a}}$ with $a_1=a_2=a_3=1$, 
the largest eigenvalue $\Omega_+^{(2)}(\lambda)$ of Eq.~(\ref{omega+}) is given in Theorem~\ref{lemma2}. 
  \begin{theorem}
  \label{lemma2}
  $\Omega_+^{(2)}(\lambda)$ is given by $x/4$, where $x$ is the maximum solution of the following equality for $x$,
\begin{align}
  &-32+64\lambda-32\lambda^2+2\lambda^3+(32-40\lambda+9\lambda^2)x\notag\\
  &+(6\lambda-10)x^2+x^3=0.
  \end{align}
  \end{theorem}

  Next, we derive $\Omega_-^{(2)}(\lambda)$ that is the largest eigenvalue of the operator defined in Eq.~(\ref{omega-}).
  In this case, since $\U{wt}({\bm{a}})=\nu-1=1$, we need to find $\bm{a}$ with $\U{wt}(\bm{a})=1$
  that achieves the largest eigenvalue of the operator $\hat{\Pi}^{\U{(ph)}}_{\bm{a}}-\lambda\hat{\Pi}$.
  For this, we again note that 
  the previous analysis~\cite{Kiyoshi2012dps} relies on a numerical method that compares each of
  all the largest eigenvalues of Eq.~(\ref{omega-}) with different $\bm{a}$ of $\U{wt}(\bm{a})=1$.
  On the other hand, we do not rely on such a numerical method and derive $\Omega_-^{(2)}(\lambda)$ in a closed form as shown 
  in Theorem~\ref{lemma3} that can be applied for any $L$ of $L\geq3$ (see Appendix~\ref{sec:apC} for the proof). 

 \begin{theorem}
  \label{lemma3}
  Among various $\bm{a}$ with $\U{wt}({\bm{a}})=1$, the largest eigenvalue of $\hat{\Pi}^{\U{(ph)}}_{\bm{a}}-\lambda\hat{\Pi}$
  realizes when $\bm{a}=0100...0$. 
 \end{theorem}
 Thanks to this Theorem, we have that $\Omega_-^{(2)}(\lambda)$ is the largest eigenvalue of the
 operator $\hat{\Pi}^{\U{(ph)}}_{\bm{a}:\bm{a}=0100...0}-\lambda\hat{\Pi}$. 
 
  Whether $\Omega_+^{(2)}(\lambda)$ is larger than $\Omega_-^{(2)}(\lambda)$ or not depends on $\lambda$.
  With a constant $\tilde{\lambda}$, which is solely dependent on $L$, $\Omega_+^{(2)}(\lambda)\leq\Omega_-^{(2)}(\lambda)$
  for $\lambda\geq\tilde{\lambda}$ and $\Omega_+^{(2)}(\lambda)>\Omega_-^{(2)}(\lambda)$ for $\lambda<\tilde{\lambda}$.
  As a result, the boundary of $(e^{(\U{b},2)},e^{(\U{ph},2)})$ consists of two convex curves determined from $\Omega_\pm^{(2)}(\lambda)$
  and a straight line with the slope $\tilde{\lambda}$ connecting them, which is shown in Fig.~\ref{fig:two}
  (see the blue curve).
  We also show the previous result of the relation $(e^{(\U{b},2)},e^{(\U{ph},2)})$ in~\cite{Kiyoshi2012dps} (see the red curve), 
  and we find that the resulting relation is almost the same. 
\\

Recall that the secret keys are generated from the $\nu=$0, 1 and 2-photon emission events,
and we take a worst case scenario that if Alice emits more than two photons, Eve has perfect knowledge on the sifted keys. 
From this consideration, the amount of privacy amplification in Eq.~(\ref{PAnu}) is upper bounded by
\begin{align}
  Qf_{\U{PA}}&=\sum^{\infty}_{\nu=0}Q^{(\nu)}h(e^{\U{(ph,\nu)}})
  \leq \sum^{\infty}_{\nu=0}Q^{(\nu)}[\gamma e^{\U{(\U{b},\nu)}}+\Omega^{(\nu)}_h(\gamma)]\notag\\
  &\leq \gamma e^{(\U{b})}+\sum^2_{\nu=0}Q^{(\nu)}\Omega^{(\nu)}_h(\gamma)+\sum_{\nu\geq3}Q^{(\nu)}\notag\\
  &= \gamma e^{(\U{b})}+Q+\sum^2_{\nu=0}Q^{(\nu)}(\Omega^{(\nu)}_h(\gamma)-1),
  \label{PAEXP}
\end{align}
where in the first inequality, we bound the convex regions of ($e^{(\U{b},\nu)},h(e^{(\U{ph},\nu)}$))
specified by a set of linear inequalities with $0<\gamma<\infty$,
\begin{align}
h(e^{(\U{ph},\nu)})\leq \gamma e^{(\U{b},\nu)}+\Omega^{(\nu)}_h(\gamma).
\end{align}
Here, $\Omega^{(\nu)}_h(\gamma)$ is the quantity depending on $\gamma$ and $\Omega^{(\nu)}(\gamma)$. 
Also, in the second inequality of Eq.~(\ref{PAEXP}),
we use the definition of the bit error rate $e^{(\U{b})}=\sum_{\nu}Q^{(\nu)}e^{\U{(\U{b},\nu)}}$. 
Since we can choose arbitrary $\gamma$ to lower-bound the amount of privacy amplification,
the lower bound on the asymptotic key generation rate in Eq.~(\ref{keyrate}) is written as
\begin{align}
  &G\geq\notag\\
  &\frac{1}{L}
  \left\{\sum^2_{\nu=0}Q^{(\nu)}-Qf_{\U{EC}}(e^{(\U{b})})-\inf_{\gamma}[\gamma e^{(\U{b})}+\sum^2_{\nu=0}Q^{(\nu)}\Omega^{(\nu)}_h(\gamma)]\right\}.
\end{align}
Here, $Q^{(\nu)}$ is chosen to maximize Eq.~(\ref{PAEXP}) as~\cite{Kiyoshi2012dps}
\begin{align}
  Q^{(\nu)}=
  \begin{cases}
    p_{\nu}&  (\nu\geq \nu_{\min}+1)\\
    Q-(1-\sum^{\nu_{\min}}_{\nu'=0}p_{\nu'})& (\nu=\nu_{\min})\\
    0 & (\nu\leq \nu_{\min}-1),
    \end{cases}
  \end{align}
where $\{p_{\nu}\}$ is the Poisson distribution with mean $L\alpha^2$
\begin{align}
p_{\nu}=e^{-L\alpha^2}(L\alpha^2)^\nu/\nu!
  \end{align}
and $\nu_{\min}$ is the integer satisfying
\begin{align}
1-\sum^{\nu_{\min}}_{\nu'=0}p_{\nu'}< Q\leq 1-\sum^{\nu_{\min}-1}_{\nu'=0}p_{\nu'}.
  \end{align}

\section{Key generation rates}
\label{sec:key}

\begin{figure}[t]
\includegraphics[width=7cm]{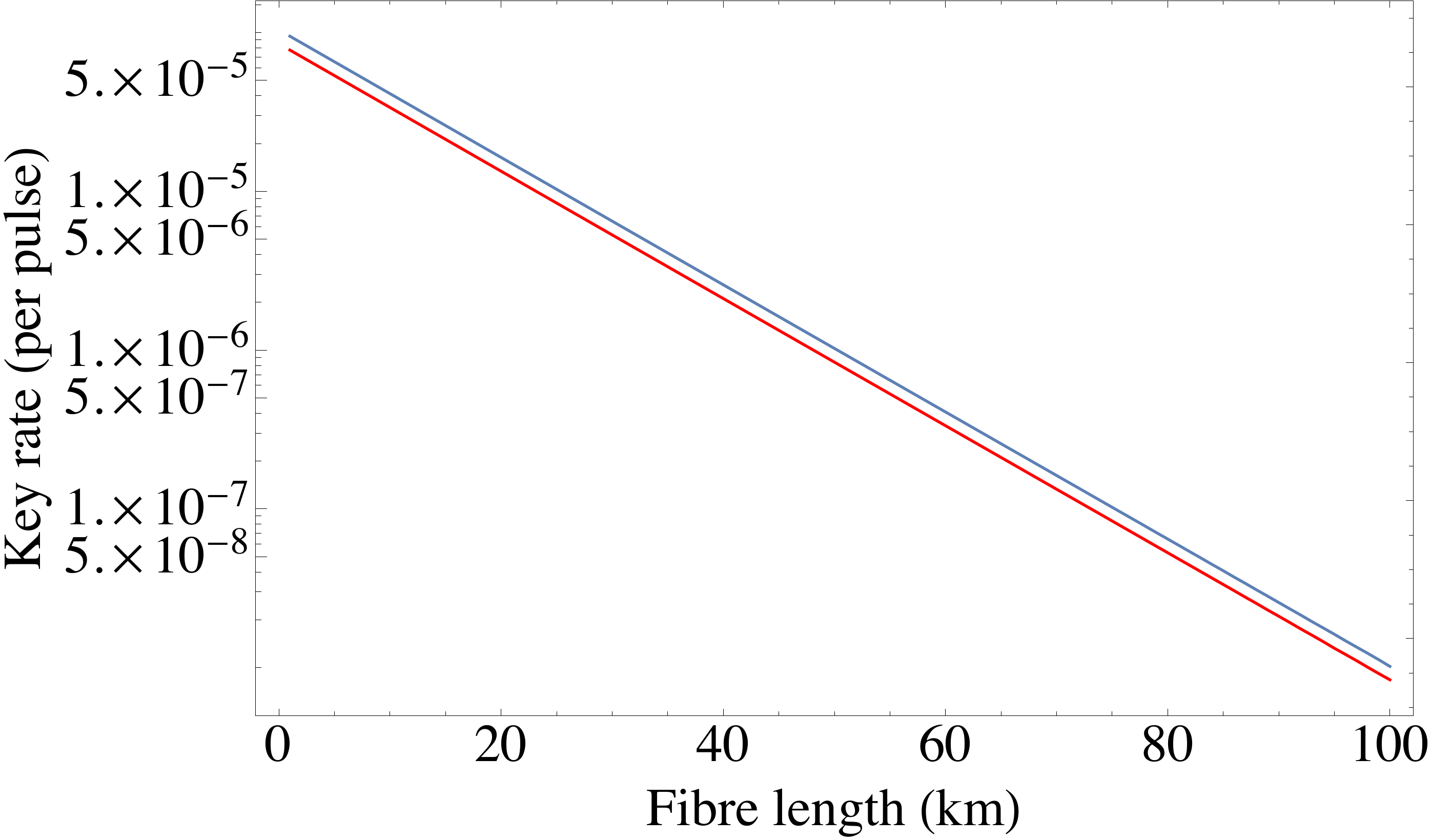}
\caption{
  Color online: The key generation rate per sending pulse of the DPS protocol based on our security analysis (blue curve)
  and the previous analysis (red curve). In these plots, we set $L=10$ and $e^{(\U{b})}=2\%$. 
}
\label{fig:key}
\end{figure}
Here, we show the key generation rates of the DPS protocol based on our analysis and the previous analysis in~\cite{Kiyoshi2012dps}.
For both cases, we suppose that the total detection probability is assumed to be 
given by $Q=(L-1)\eta\alpha^2e^{-(L+1)\eta\alpha^2}$, where
we assume that the transmittance of the channel including the detection probability is given by $\eta=0.1\times 10^{-0.2l/10}$, and
for simplicity we adopt $f_{\U{EC}}(e^{(\U{b})})=h(e^{(\U{b})})$. 
In Fig.~\ref{fig:key}, we show the key rates for the case of
$e^{(\U{b})}=2\%$ and $L=10$ by optimizing the mean photon number $\alpha^2$.
As a result, we find that the secure key rate of the DPS protocol based on our analysis is 1.22 times as high as the
previous one. 

We note that the two key generation rates in Fig.~\ref{fig:key} do not decrease drastically after a certain distance.
This is so because we consider the constant bit error rate, which is independent of the distance.

\section{conclusion}
\label{sec:conc}
In conclusion, we have proven the information-theoretic security proof for the DPS QKD protocol based on 
the complementarity approach. 
As a result, we found that our security proof provides a slightly better key generation rate compared to the
previous security proof based on the Shor-Preskill approach~\cite{Kiyoshi2012dps}.
This improvement is obtained since the complementarity approach can incorporate more detailed information on Alice's sending state
to estimate the leaked information to Eve. 
In particular, we have employed the information 
that the intensity of Alice's sending states is weak, and thanks to this additional information we have obtained 
tighter upper bounds on the leaked information to Eve compared to those in the previous proof~\cite{Kiyoshi2012dps}.
Moreover, we have removed the necessity of the numerical calculation, which was needed in~\cite{Kiyoshi2012dps}  
to derive the leaked information on the two-photon emission events.
This leads to an advantage that our security proof
enables us to evaluate the security of the DPS protocol with any block size $L$ of $L\geq3$.

\section*{Acknowledgments}
G. K. and K. T. acknowledge support from CREST, JST. 
Y. T. acknowledges support from Grant-in-Aid for JSPS Fellows (KAKENHI Grant No. JP17J03503) and
the support from the Program for Leading Graduate Schools: Interactive Materials Science Cadet Program.
T. S. acknowledges the support from the ImPACT Program of Council for Science, Technology and Innovation
(Cabinet Office, Government of Japan).
AM acknowledges support from Grant-in-Aid for JSPS Fellows (KAKENHI Grant No. JP17J04177).

\appendix

\begin{widetext}

  \section{Proof of Eq.~(\ref{unieph})}
  \label{sec:apA}
      Here, we prove Eq.~(\ref{unieph}). First, from the definition of $\hat{U}$ in Eq.~(\ref{unitary}), we have
      \begin{align}
        \hat{U}\hat{P}(\ket{s}_{A,i})\hat{P}(\ket{i'}_B)\hat{U}^{\dagger}
        =\hat{P}(\ket{s\oplus \delta_{i,i'}}_{A,i})\hat{P}(\ket{i'}_{B}).
        \label{unitaryph}
      \end{align}
      By using Eqs.~(\ref{ephj}) and (\ref{unitaryph}), a direct calculation leads to the following equation
      \begin{align}
  &\hat{U}\hat{e}^{(\U{ph})}_{j}\hat{U}^{\dagger}=\hat{e}^{(\U{ph})}_{j}\notag\\
        &=\sum_{{\bm{a}}}\hat{P}(\ket{{\bm{a}}}_A)\otimes\Big[
          \delta_{a_j,0}\delta_{a_{j+1},1}\kappa_{j}\hat{P}(\ket{j}_B)
          +\delta_{a_j,1}\delta_{a_{j+1},0}\kappa_{j+1}\hat{P}(\ket{j+1}_B)
          +\delta_{a_j,1}\delta_{a_{j+1},1}\Big(\kappa_{j+1}\hat{P}(\ket{j+1}_B)+\kappa_{j}\hat{P}(\ket{j}_B)\Big)
          \Big]\notag\\
        &=\sum_{{\bm{a}}}\hat{P}(\ket{{\bm{a}}}_A)\otimes
  \Big[\kappa_j
    \delta_{a_{j+1},1}\hat{P}(\ket{j}_B)+\kappa_{j+1}\delta_{a_j,1}\hat{P}(\ket{j+1}_B)\Big],
      \end{align}
      which concludes Eq.~(\ref{unieph}).
      Note that in the final equation, we use the property of the Kronecker delta 
      $\delta_{a_{j},0}+\delta_{a_{j},1}=1$. 

      \section{Proof of Theorem~\ref{theorem1}}
      \label{sec:appTheorem}
      In this appendix, we prove Theorem~\ref{theorem1}. For this, we consider bounding the convex achievable 
    reagion ($e^{(\U{b},1)},e^{(\U{ph},1)}$) by a straight line with 
    the slope either $\lambda=3+\sqrt{5}=:\lambda_0$ or $\lambda=\lambda'(<\lambda_0)$.
    Here, from Eq.~(\ref{omega1}), $\lambda>\lambda_0$ never appears in the discussion since it is trivially understood
    from Eqs.~(\ref{relationeph}) and (\ref{omega1}) that $\lambda=\lambda_0$ gives a tighter bound on $e^{(\U{ph},1)}$ 
    than $\lambda>\lambda_0$. In a lower bit error rate regime, the convex achievable region is tightly bounded with
    $\lambda=\lambda_0$.
    On the other hand, in a higher bit error rate regime, $\lambda'$ gives a tighter bound. 
    Hence, we derive the threshold bit error rate $e^{*(\U{b},1)}$ that for $e^{(\U{b},1)}\leq (>)e^{*(\U{b},1)}$, the convex achievable
    region is tightly bounded with the slope $\lambda=\lambda_0~(\lambda')$. 
    This threshold can be derived by solving the following equation:
    \begin{align}
      \lambda_0e^{(\U{b},1)}+\Omega^{(1)}_+(\lambda_0)=\lambda' e^{(\U{b},1)}+\Omega^{(1)}_+(\lambda').\label{thres}
    \end{align}
    By using Lemma~\ref{lemma1}, Eq.~(\ref{thres}) leads to 
    \begin{align}
      e^{(\U{b},1)}=\frac{(3-\sqrt{5}-\lambda')}{2(3-2\lambda'-\sqrt{1+2\lambda'^2})}=:f(\lambda').
    \end{align}
    By taking a limit of $\lambda'\rightarrow\lambda_0$, we can derive the threshold bit error rate as 
    $e^{*(\U{b},1)}=\lim_{\lambda'\rightarrow\lambda_0} f(\lambda')=(10-3\sqrt{5})/22$. 
    This means that for $0\leq e^{(\U{b},1)}\leq e^{*(\U{b},1)}$, $\lambda=\lambda_0$ is the optimal slope to bound 
    $e^{(\U{ph},1)}$, and for $e^{(\U{b},1)}>e^{*(\U{b},1)}$, the optimal slope $\lambda$
    (with $0<\lambda< \lambda_0$) changes according to the bit error rate as shown in Eq.~(\ref{nu1high}). 
      
    \section{Proof of Theorem~\ref{lemma3}} 
    \label{sec:apC}
    In this appendix, we prove Theorem~\ref{lemma3} in the main text.
    We first prove the Theorem for the case of $L=3$ and 4, and after that we prove it for the case of $L\geq5$.
    
    For the cases of $L=3$ and 4, 
    through a direct comparison of the largest eigenvalues of the operator $\hat{\Pi}^{\U{(ph)}}_{\bm{a}}-\lambda\hat{\Pi}$
    among various $\bm{a}$ with $\U{wt}(\bm{a})=1$, it is easy to confirm that the operator gives the largest eigenvalues 
    when $a_2=1$. 
    
    Next, from now on, we move on to the general case of $L\geq5$.
    We first introduce two functions $F(L,x,w,y)$ and $g_s(L,x,w,m)$ and investigate their properties,
    which we use in the main part of the proof in Appendix~\ref{appCmatrix}. 
    Throughout the discussion below, we assume $L\geq5$, $0\leq x$, $-1\leq y\leq 1$ and $0<w<\infty$. 
    
    \subsection{Function $F(L,x,w,y)$}
    First, we introduce $F(L,x,w,y)$, which is defined as
    \begin{align}
  F(L,x,w,y)=&\frac{1}{2}\cosh{Lx}-2w\cosh(L-1)x+\left(2w^2-\frac{1}{2}\right)\cosh(L-2)x+2w^2\cosh(L-4)x\notag\\
  &+2w(2w\cosh{x}-\cosh{2x})\cosh{(L-3)xy}.
  \label{funcF}
    \end{align}
From this definition, it is easy to confirm that 
    \begin{align}
      F\left(L,x,\frac{\cosh{2x}}{2\cosh{x}},y\right)=-\sinh x\sinh(L-5)x\leq0
      \label{proF1}
  \end{align}
    and for $0\leq w\leq 1/2$, 
        \begin{align}
          F(L,0,w,y)=4w(2w-1)\leq0.
                  \label{proF2}
        \end{align}
        Here, $w=\frac{\cosh{2x}}{2\cosh{x}}$ is a monotonically increasing and one-to-one function
        from the domain $\mathbb{R}_{\geq 0}$ to $\mathbb{R}_{\geq 1/2}$.
        Hence, an inverse function $f(w)$ can be defined from $\mathbb{R}_{\geq 1/2}$ to $\mathbb{R}_{\geq 0}$ as
        \begin{align}
          x_w=
          \begin{cases}
            0&\U{if}~w\leq 1/2\\
            f(w)&\U{if}~w> 1/2. 
            \end{cases}
            \end{align}
With this notation and Eqs.~(\ref{proF1}) and (\ref{proF2}), we have that
        \begin{align}
          L>5~\U{and}~w\neq\frac{1}{2}\Rightarrow F(L,x_w,w,y)<0,\label{P1}\\
                    L=5~\U{or}~w=\frac{1}{2}\Rightarrow F(L,x_w,w,y)=0.
          \label{P2}
        \end{align}
        Also, we have a property
        \begin{align}
          \lim_{x\to+\infty}F(L,x,w,y)=+\infty.
          \label{P3}
          \end{align}
        From Eqs.~(\ref{P1})-(\ref{P3}) and the continuity of $F(L,x,w,y)$, we obtain
        \begin{align}
\forall L, w, y, \exists x\geq x_w,~F(L,x,w,y)=0. 
\end{align}
        From this equation, we can define the following function
        \begin{align}
          x_{\max}(L,w,y)=\max\{x|F(L,x,w,y)=0\}.
          \label{xmax}
          \end{align}
        In the case of $L>5~\U{and}~w\neq\frac{1}{2}$, we can confirm from (\ref{P1}) that $x_{\max}(L,w,y)$ is strictly larger than
        $x_w$: 
        \begin{align}
          x_w<x_{\max}(L,w,y).
          \label{xw<xmax}
        \end{align}
          Also, Eq.~(\ref{xw<xmax}) holds in the case of $w=\frac{1}{2}$ or $L=5$. 
        For $w=\frac{1}{2}$ and $L\ge5$, Eq.~(\ref{xw<xmax}) is satisfied since
        \begin{align}
          \frac{d}{dx}F\left(L,x,\frac{1}{2},y\right)|_{x=0}=0~\U{and}~
          \frac{d^2}{dx^2}F\left(L,x,\frac{1}{2},y\right)|_{x=0}=-2(L-2)<0.
          \end{align}
Also, for $L=5$ and $w\neq1/2$, Eq.~(\ref{xw<xmax}) is satisfied since
        \begin{align}
          \frac{d}{dx}F(5,x,w,y)|_{w=\frac{\cosh{2x}}{2\cosh{x}}}=-\frac{1}{2}\cosh^{-2}x(1+\cosh{2x}\cosh{2xy})(3\sinh{x}+\sinh{3x})<0
\end{align}
        holds.
        
Next, since
\begin{align}
\forall x\geq x_w,~2w\cosh{x}-\cosh{2x}\leq 0
\end{align}
holds, we have the following relation if $0\leq y\leq1$
\begin{align}
\forall x\geq x_w,~\frac{d}{dy}F(L,x,w,y)=2w(2w\cosh{x}-\cosh{2x})(L-3)x\sinh{(L-3)xy}\leq0.
  \end{align}
From this equation and Eq.~(\ref{xw<xmax}), we obtain
\begin{align}
F(L,x_{\max}(L,w,y),w,y)\geq F(L,x_{\max}(L,w,y),w,1),
\end{align}
which leads to the inequality
\begin{align}
x_{\max}(L,w,y)\le x_{\max}(L,w,1)
  \label{max:y1}
\end{align}
if $0\leq y\leq1$. 
As for $-1\leq y\leq 0$, from the definitions of the function $F(L,x,w,y)$ and
$x_{\max}$, we find that these are even functions of $y$.
Therefore, for $-1\leq y\leq1$,
\begin{align}
x_{\U{max}}(L,w,y)=x_{\U{max}}(L,w,|y|)\le x_{\max}(L,w,1)
\end{align}
is satisfied.

\subsection{Property of the function $g_s(L,x,w,m)$}
Next, we introduce the function $g_s(L,x,w,m)$ for $s\in\{-1,1\}$, which is given by
\begin{align}
  g_s(L,x,w,m)=\cosh\left(\frac{L-1}{2}+sm\right)x-2w\cosh\left(\frac{L-3}{2}+sm\right)x.
  \label{gs}
\end{align}
This function has a following property that we use in Appendix~\ref{appCmatrix}. 
\begin{property}
  \label{progs}
For $|m|\leq \frac{L-5}{2}$, 
  \begin{align}
    g_s\left(L,x_{\max}\left(L,w,\frac{2m}{L-3}\right),w,m\right)>0
    \label{progs1}
  \end{align}
and for $|m|=\frac{L-3}{2}$, 
  \begin{align}
    g_1(L,x_{\max}(L,w,y),w,m)\neq0~\U{or}~g_{-1}(L,x_{\max}(L,w,y),w,m)\neq0.
        \label{progs2}
  \end{align}
\end{property}
    {\it Proof of Property~\ref{progs}}

    First, we prove Eq.~(\ref{progs1}). 
    Since $g_s(L,x,w,m)$ is a monotonically-decreasing function for $w$, we have
\begin{align}
  x>x_w,|m|\leq\frac{L-5}{2}\Rightarrow g_s(L,x,w,m)>
  g_s\left(L,x,\frac{\cosh{2x}}{2\cosh{x}},m\right)=\sinh\left(\frac{L-5}{2}+sm\right)x\tanh{x}\geq0,
  \label{gs>0}
  \end{align}
where we use the fact $x>x_w\Rightarrow \frac{\cosh{2x}}{2\cosh{x}}>w$.
By combining Eqs.~(\ref{gs>0}) and (\ref{xw<xmax}), we obtain 
\begin{align}
g_s\left(L,x_{\max}\left(L,w,\frac{2m}{L-3}\right),w,m\right)>0
\end{align}
for $|m|\leq\frac{L-5}{2}$, which concludes Eq.~(\ref{progs1}).

As for the proof of Eq.~(\ref{progs2}), we use 
\begin{align}
  &g_1\left(L,x,w,\frac{L-3}{2}\right)-g_{-1}\left(L,x,w,\frac{L-3}{2}\right)\cosh(L-3)x\notag\\
  =&g_{-1}\left(L,x,w,-\frac{L-3}{2}\right)-g_1\left(L,x,w,-\frac{L-3}{2}\right)\cosh(L-3)x\notag\\
  =&\sinh{x}\sinh{(L-3)x}>0,
\end{align}
for $x>0$ and since $\cosh(L-3)x>0$, we can conclude Eq.~(\ref{progs2}).

\subsection{Calculation of $\hat{\Pi}^{(\U{ph})}_{{\bm{a}}}-\lambda\hat{\Pi}$}
\label{appCmatrix}    
    Next, we calculate the matrix elements of $\hat{\Pi}^{(\U{ph})}_{{\bm{a}}}-\lambda\hat{\Pi}
      =(\hat{\Pi}^{(\U{ph})}_{{\bm{a}}}-\lambda\hat{\Pi}|_{j,k})_{j,k}$, which is a tridiagonal symmetric matrix.
            For convenience of notation, we denote the index of low and column of the matrix by 
      $j,k\in\{-\frac{L-1}{2},-\frac{L-3}{2},...,\frac{L-3}{2},\frac{L-1}{2}\}$ (not by $\{1,2,...,L-1,L\}$). 
      The diagonal elements are given by
      \begin{align}
        \hat{\Pi}^{(\U{ph})}_{{\bm{a}}}-\lambda\hat{\Pi}|_{j,j}
=
        \begin{cases}
          \delta_{a_{\pm\frac{L-3}{2}},1}-\lambda/2&(j=\pm\frac{L-1}{2})\\
          (\delta_{a_{j-1},1}+\delta_{a_{j+1},1})/2-\lambda/2&(j\neq\pm\frac{L-1}{2}),
                 \label{digA}
          \end{cases}
        \end{align}
and the off-diagonal elements are given by
      \begin{align}
        \hat{\Pi}^{(\U{ph})}_{{\bm{a}}}-\lambda\hat{\Pi}|_{j,j-1}=
        \begin{cases}
          \sqrt{2}\lambda/4&(j=\frac{L-1}{2})\\
          \lambda/4&(-\frac{L-5}{2}\leq j\leq\frac{L-3}{2})\\
                 \sqrt{2}\lambda/4&(j=-\frac{L-3}{2}).
          \label{offA}
        \end{cases}
        \end{align}
      Since we now consider $\U{wt}(\bm{a})=\nu-1=1$, we need to choose only one $j$ such that $a_j=1$. 
      If we define $m$ that satisfies $a_m=1$,      
      $\hat{\Pi}^{(\U{ph})}_{{\bm{a}}}-\lambda\hat{\Pi}$ is characterized by $m$, and we define 
      $\hat{A}^{(m)}:=\hat{\Pi}^{(\U{ph})}_{{\bm{a}}:a_m=1}-\lambda\hat{\Pi}$.
      Below, we classify $m$ into three cases, and for each of all the cases we calculate the matrix elements of
      $\hat{A}^{(m)}=(\hat{A}^{(m)}|_{j,k})_{j,k}$. 

      (I) For $|m|=\frac{L-1}{2}$, 
\begin{align}
  \hat{A}^{(s\frac{L-1}{2})}|_{j,k}=\frac{\delta_{j,s\frac{L-3}{2}}-\lambda}{2}\delta_{j,k}+
  \lambda\frac{1+(\sqrt{2}-1)\delta_{|j+k|,L-2}}{4}\delta_{|j-k|,1},
  \label{caseI}
\end{align}  
where $s\in\{-1,1\}$. 

(II) For $|m|=\frac{L-3}{2}$, 
\begin{align}
  \hat{A}^{(s\frac{L-3}{2})}|_{j,k}=
  \frac{2\delta_{j,s\frac{L-1}{2}}+\delta_{j,s\frac{L-5}{2}}-\lambda}{2}\delta_{j,k}+
  \lambda\frac{1+(\sqrt{2}-1)\delta_{|j+k|,L-2}}{4}\delta_{|j-k|,1}.
    \label{caseII}
\end{align}  

(III) For $|m|\leq \frac{L-5}{2}$, 
\begin{align}
  \hat{A}^{(m)}|_{j,k}=
  \frac{\delta_{j,m-1}+\delta_{j,m+1}-\lambda}{2}\delta_{j,k}+
  \lambda\frac{1+(\sqrt{2}-1)\delta_{|j+k|,L-2}}{4}\delta_{|j-k|,1}.
    \label{caseIII}
\end{align}  
Now, we prove Theorem~\ref{lemma3} that the largest eigenvalue of $\hat{A}^{(m)}$ realizes when $|m|=\frac{L-3}{2}$ 
for $L\geq5$ and $0<\lambda<\infty$. 
First, from the above expressions in Eqs.~(\ref{caseI}) and (\ref{caseIII}),
$\hat{A}^{(\pm\frac{L-1}{2})}\leq \hat{A}^{(\pm\frac{L-5}{2})}$ are obtained, and 
hence we need not to consider the case (I) for deriving the largest eigenvalue of $\hat{A}^{(m)}$. 

Next, we compare the eigenvalues of the cases (II) and (III). 
For this, we prepare the vector $v^{(m)}=(v^{(m)}|_j)_j$ that has $L$ elements. 
As well as the matrix representation, we denote the index of column of the vector by 
$j\in\{-\frac{L-1}{2},-\frac{L-3}{2},...,\frac{L-3}{2},\frac{L-1}{2}\}$.
For $s\in\{-1,1\}$, $j_-\in\{-\frac{L-3}{2},-\frac{L-5}{2},...,m-2,m-1\}$ and $j_+\in\{m+1,m+2,...,\frac{L-5}{2}\frac{L-3}{2}\}$, 
we define $v^{(m)}$ as
\begin{align}
  v^{(m)}|_{s\frac{L-1}{2}}=\frac{1}{\sqrt{2}}g_s(L,x,\lambda^{-1},m),\\
  v^{(m)}|_m=g_1(L,x,\lambda^{-1},m)g_{-1}(L,x,\lambda^{-1},m),\\
  v^{(m)}|_{j_-}=g_{-1}(L,x,\lambda^{-1},m)\cosh\left(\frac{L-1}{2}+j_-\right)x,\\
  v^{(m)}|_{j_+}=g_1(L,x,\lambda^{-1},m)\cosh\left(\frac{L-1}{2}-j_+\right)x,
\end{align}
where $x\in\mathbb{R}$ is a free parameter.
In this case, the following equation holds.
\begin{align}
  \left[\hat{A}^{(m)}-\frac{\lambda(\cosh x-1)}{2}\hat{I}\right]\cdot v^{(m)}|_j=
  -\frac{\lambda}{4}\delta_{j,m}F\left(L,x,\lambda^{-1},\frac{2m}{L-3}\right).
  \label{charac}
\end{align}
Here we recall that $g_{s}(L,x,\lambda^{-1},m)$ and $F(L,x,\lambda^{-1},\frac{2m}{L-3})$ are defined in Eqs.~(\ref{gs}) and
(\ref{funcF}), respectively. 
In the following, we set $x=x_{\max}(L,\lambda^{-1},\frac{2m}{L-3})$.
From its definition in Eq.~(\ref{xmax}), we have that the rhs of Eq.~(\ref{charac}) is zero, which leads to
\begin{align}
  \left[\hat{A}^{(m)}-\frac{\lambda(\cosh x_{\max}(L,\lambda^{-1},\frac{2m}{L-3})-1)}{2}\hat{I}\right]\cdot v^{(m)}|_j=0.
\end{align}
Moreover, thanks to Property~\ref{progs}, we find that all the elements of $v^{(m)}$ never become zero simultaneously, and
therefore we conclude that $v^{(m)}$ and $\frac{\lambda}{2}(\cosh x_{\max}(L,\lambda^{-1},\frac{2m}{L-3})-1)$
represent the eigenvector and the eigenvalue of $\hat{A}^{(m)}$, respectively. 
In particular, if $|m|\leq\frac{L-5}{2}$,
all the elements of $v^{(m)}$ are guaranteed to be positive due to Eq.~(\ref{progs1}) in Property~\ref{progs}. 
By combining this fact and the fact that $\hat{A}^{(m)}$ is an Hermite operator with non-negarive off-diagonal elements, 
we conclude that the eigenvalue $\frac{\lambda}{2}(\cosh x_{\max}(L,\lambda^{-1},\frac{2m}{L-3})-1)$
is the largest eigenvalue of $\hat{A}^{(m)}$ if $|m|\leq\frac{L-5}{2}$. 
Finally, from Eq.~(\ref{max:y1}), we have that
\begin{align}
  \frac{\lambda}{2}\left[\cosh x_{\max}\left(L,\lambda^{-1},\frac{2m}{L-3}\right)-1\right]
  \leq \frac{\lambda}{2}(\cosh x_{\max}(L,\lambda^{-1},1)-1). 
  \end{align}
Since the rhs represents the eigenvalue of $\hat{A}^{(\frac{L-3}{2})}$, we conclude that
$\hat{A}^{(\frac{L-3}{2})}$ has an eigenvalue that is no smaller than the largest eigenvalue of $\hat{A}^{(m)}$ with 
$|m|\leq \frac{L-5}{2}$.

\end{widetext}

\end{document}